\documentclass[]{aa}
\usepackage{times}
\usepackage{mathptm}
\usepackage{psfig}
\usepackage{german}
\usepackage{graphics}
\begin{document}

\thesaurus{08(08.09.2; 08.03.2; 08.03.4; 08.13.1)}
\title{Wind circulation in selected rotating magnetic early-B stars }

\author{Myron A. Smith \inst{1} \and Detlef Groote \inst{2} }
\institute{Space Telescope Science Institute/CSC, 3700 
San Martin Dr. Baltimore, MD 21218 and Catholic University of America, 
Washington, D.C.
\and Hamburger Sternwarte, Gojenbergsweg 112, 21029 Hamburg, Germany
}
\offprints{Myron A. Smith}
\mail{msmith@stsci.edu}
\date{received date; accepted date}
\maketitle

\begin{abstract}

  The rotating magnetic B stars are a class of variables 
consisting of He-strong and some $\beta$\,Cep stars which have
oblique dipolar magnetic fields. Such stars
develop co-rotating, torus-shaped clouds by channeling wind particles 
from their magnetic poles to circumstellar regions centered around the
plane of their magnetic equators. The rotation of the cloud-star complex 
permits the study of absorptions from the cloud as it occults the star. 
In this paper we describe a quantitative analysis of archival {\it IUE} 
data to map the properties of these clouds over four stars (HD\,184927, 
$\sigma$\,Ori\,E, $\beta$\,Cep, and HR\,6684). By computing spectral synthesis 
models for these stars, we find that only $\beta$\,Cep has a solar-like 
metallicity. Our analysis also shows that the metal composition 
across the surfaces of all these stars is at least approximately homogeneous.

  Using the Hubeny code {\it CIRCUS}, we demonstrate that the periodic 
variations of broad-band ultraviolet continuum fluxes can be explained fully by
the absorptions of the co-rotating cloud. We show next that among 
selected lines, those arising from low-excitation states are selectively 
affected by cloud absorption and turbulence. Our analysis also quantifies 
the cloud temperatures and column densities required to match the 
absorptions of a number of weak to moderate strength resonance lines. 
These temperatures increase with the ionization potential of the parent 
ions of these various 
lines, a result which is consistent with radiative equilibrium models in 
which temperature increases with proximity to the star's surface. 
Although these attributes appear stable from one epoch to another, dynamic 
processes are nonetheless at work. Both the strengths and widths of resonance 
lines at occultation phases indicate the presence of a turbulence in the cloud 
which increases inwards.  

 The spectroscopic hallmark of this stellar class is the presence of
strong C\,IV and N\,V resonance line absorptions at occultation phases and 
of redshifted emissions of these lines at magnetic pole-on phases. The 
emissions have characteristics which seem most compatible with their 
generation by high-energy shocks at the wind-cloud interface, as predicted 
recently by Babel (\cite{bab}).

\end{abstract}

\section{Introduction}
\label{intro} 

  The He-strong Bp stars are a subclass of early-type B stars with an array 
of remarkable properties, chief among which are their oblique dipolar 
magnetic fields, magnetically controlled winds, 
and chemical surface anomalies (e.g. Bohlender et al. \cite{bohal1}, 
Bohlender \cite{boh1}, Bolton \cite{bol1}, Shore \cite{shoal2}). 
These conditions provide a rich
testbed for the study of magnetic and hydrodynamical processes on these stars 
and conceivably on stars of related active O- and B-subclasses in which wind 
and photospheric line variability has so far gone undiagnosed.
Several seminal papers on the He-strong star $\sigma$\,Ori\,E have 
demonstrated that several peculiarities in the ultraviolet spectrum
of this star can be explained
by circumstellar clouds forced into co-rotation over the magnetic equator.
For example,  Groote \& Hunger (\cite{grohun1}) showed that high-level Balmer 
lines appear at certain rotational phases, demonstrating the presence of an 
occulting, comparatively low-density gas. This circumstellar medium can also be 
observed as periodically modulated H$\alpha$ emissions (e.g., Short \& Bolton 
\cite{shobol}).

  Using spectra from the {\it International Ultraviolet Explorer
(IUE)} satellite, Shore and collaborators (Shore \& Brown 
\cite{shobro} (SB90), Shore et al. \cite{shoal1}) demonstrated 
that a common property of winds in both He-strong and He-weak Bp stars 
is that 
they are channeled by a dipolar magnetic field to a torus-shaped magnetosphere 
(cloud) located over the magnetic equator. This torus may well be be located 
in a different volume than the {\em plasmasphere}
suggested by radio studies (Havnes \& Goertz \cite{havgoe}, Drake et al. 
\cite{draal1} and Drake \cite{dra1}), 
but it is likely to be related. Following the initial ideas of Shore 
(\cite{sho}), Babel \& Montmerle (\cite{babmon2}) and Babel (\cite{bab}) 
have discussed a model in which shocks are excited at the wind-torus 
interface, creating high energy emission. The magnetic and rotational axes 
of these stars are separated by seemingly arbitrary angles. In most cases 
this angle is large enough to provide an opportunity to observe the 
time-variability of emission and absorption components of the strong lines 
and thus to study the azimuthal structure of the clouds as the wind-cloud 
complex co-rotates. While these variations are often large and complex,
all indications are that they are rather stable over
decades (e.g., SB90, Bolton \cite{bol1}, Henrichs et al. \cite{henal2}).
 
  Henrichs et al. (\cite{henal1}) recently proposed the existence of a 
distinct new class of {\em magnetic rotators} (hereinafter, {\em 
rotating magnetic B stars}). The three stars so far assigned to this 
group, $\beta$\,Cep (HD\,205021), 
HR\,6684 (HD\,163472), and HD\,184927, are characterized 
by redshifted emissions of the C\,IV resonance lines. To this small 
class, we would add the extensively 
studied, He-strong star $\sigma$\,Ori\,E. The Si\,IV and C\,IV resonance lines
of this star also show redshifted emissions at phases of magnetic extrema 
(Bohlender et al. \cite{bolal1}). We have examined the sequences of
C\,IV and 
N\,V resonance profiles in three He-strong stars in the SB90 sample with 
known rotational periods (HD\,37017, HD\,37776, HD\,64740) and find that 
each exhibits phase-correlated redshifted emissions. An examination of {\it 
NEWSIPS}-processed spectra of HD\,36385 and HD\,133518 shows that the C\,IV
resonance lines in these stars' spectra also exhibit variable, redshifted 
emissions. The periodic emissions of the N\,V and C\,IV lines of the hot O7\,V
star $\theta ^{1}$ Ori\,C
are well known. Walborn \& Nichols (\cite{walnic}) and Gagn\'e et al.
(\cite{gagal1}) have suggested that this star is a hot analog to the
Bp stars. 
Because of the similar variabilities of the red halves of their
C\,IV, N\,V profiles, it is likely that
these stars, and indeed perhaps all He-strong variables, should
be included in this new class.

  In this paper we provide the first detailed analysis of spectral lines
to determine conditions in co-rotating clouds attached to the rotating
magnetic B stars. We undertake this study by using ultraviolet lines  
obtained entirely from archival echellograms obtained with the {\em IUE} satellite.
We have three specific goals:
(1) to estimate physical parameters of clouds (temperature, density,
extent, metallicity, and turbulence), 
(2) to study the relevance of {\em Bp-like} surface chemical composition 
(metal-depletion) on cloud properties, and 
(3) to examine the processes responsible for resonance line emissions and
anomalous absorptions at certain phases.

  For these purposes we have chosen the first four stars discussed above: 
$\sigma$\,Ori\,E (the most well studied He-strong star),
HD\,184927 (a sharp-lined, magnetic He-strong star; 
Wade et al. \cite{wadal1})), $\beta$\,Cep (a composition-normal star with
magnetic wind properties similar to the He-strong stars; 
it is also the prototype of the $\beta$\,Cephei 
pulsational variables), and HR\,6684 (a $\beta$\,Cephei variable with a
moderate rotational velocity; Jerzykiewicz \cite{jer}, Kubiak \& Seggewiss
\cite{kubseg}). These last two stars are pulsators and are not He-strong stars.
The selection of these four stars was guided in part by our desire to 
include both intrinsically slow ($\beta$\,Cep) and moderately rapid rotators 
($\sigma$\,Ori\,E and HR\,6684), as well as stars observed at both low and high 
aspect angles. $\beta$\,Cep and $\sigma$\,Ori\,E are
probably intrinsically moderate and rapid rotators, respectively,
viewed nearly equator-on.
HD\,184927 is an intrinsically slow rotator viewed at low inclination. Less
is known abut HR\,6684, but its low $v$\,sin\,$i$ and possible rotation period
near 3.75 days (Henrichs et al. \cite{henal1}), suggest that it is a moderate
rotator observed at low inclination, an interpretation supported by the
single-wave variations of its magnetic and optical He\,I line variations
(Wade et al. \cite{wadal1}). In order to describe the geometry of the
magnetospheres, it is important that the  sample of stars studied has a
variety of rotational velocities and be viewed from a range of inclination 
angles.

\section{Reduction and analysis of archival IUE data }
\label{dataan}

\subsection{Data reduction}

   The {\it IUE} data for this program were the NEWSIPS-extracted,
large-aperture high dispersion spectral files from the Short Wavelength
Prime (SWP) camera. These were downloaded from the MAST
\footnote{Multi-Mission Archive at Space Telescope Science Institute, in
contract to NASA.} data archive.
The data used were all the {\it IUE} SWP/large-aperture images for
$\sigma$\,Ori\,E (29), HD\,184927 (30), $\beta$\,Cephei (31), and HR\,6684 (41),
except that for $\sigma$\,Ori\,E the images SWP\,04840 and SWP\,27259 were
excluded. The latter spectra were observed at isolated times
and thus do not lend themselves to analysis of short-term variations.
The high-dispersion SWP-camera data for both NEWSIPS
and IUESIPS are known to depend on errors in echelle {\em ripple}
correction and background subtraction. Ripple-correction errors were
avoided by working with net flux data for individual echelle orders.
To account for background
extraction errors (Smith \cite{smi}), we derived corrections first in the
$\beta$\,Cep and HD\,184927 spectra by forcing the depths of the Si\,III,
Si\,IV, C\,IV, and occasionally other resonance lines to zero.
This procedure was generally insensitive to rotation because most of the
resonance lines we studied reside well up on the flat or even damping parts of
their curves of growth, and their line cores are well broadened by saturation.
The zero-point corrections we derived were typically 5--10\% of the
continuum level.
For lines in short-wavelength orders, we interpolated corrections
from the Ly\,$\alpha$ and other nearby orders.

\subsection{Computation of simulated spectra and spectral absorptions}

  We used a suite of codes written by I. Hubeny and collaborators to
compute model line profiles to compare with the IUE data. The first code
is {\it SYNSPEC}, a spectral line synthesis code developed for input
non-LTE or line-blanketed model atmospheres by Hubeny et al.
(\cite{hubal1}). We used standard LTE atmospheres by Kurucz (\cite{kur2}) in
our work. {\it SYNSPEC} is embedded within an IDL wrapper (Hubeny \cite{hub1})
such that we could easily calculate spectra with a variety of trial atmospheric
abundances. {\it SYNSPEC} was then run to generate synthetic spectra
similar to the observed ones by computing a disk-integrated, rotationally
broadened spectrum and to convolve it to
the spectral resolution (13\,000) of high dispersion {\it IUE} spectra.

  Our principal objective in this paper is to assess the effect of
torus-shaped magnetospheric clouds on a star's spectral lines
(Shore \& Brown \cite{shobro}) as they alternatively move
onto and off the projected stellar
surface. To simulate the signatures of a cloud on the composite spectrum,
we used the Hubeny {\it CIRCUS} program (Hubeny \cite{hub1}, Hubeny \& Heap
\cite{hubhea}).
This code was written to compute line absorptions and/or emissions
of a gas cloud situated either in front or off the limb of a reference star.
{\it CIRCUS} requires the user to specify physical cloud parameters such as
temperature, density, geometry, composition, microturbulence, column depth,
and areal coverage factor. Although the code
can accommodate as many as three separate clouds
we considered only a single
homogeneous cloud in nearly all our simulations.
In its solution of the radiative transfer equation within a cloud,
{\it CIRCUS} computes line emissions and absorptions separately.
This feature can be used to compute
the cloud spectrum either in LTE (adding differential emission and absorption
terms) or in the two-level non-LTE approximation, in which only the
absorption terms are calculated.

  The code proceeds by first consulting a Kurucz (\cite{kur1}) line library of
atomic absorption parameters and computing an opacity spectrum for
a user-specified temperature and electron density.
The spectra were computed at a spacing of 0.01 Angstroms, which oversamples
the contributing lines. The optical depth in each line can then be
determined using an input column density.
The surface of the star is divided into a 100$\times$100
grid and the local intensity
spectrum is evaluated at each grid point, thereby taking into account
the effects of foreshortening and limb darkening. Typically, we
computed {\it CIRCUS} models in pairs. In the first case we assumed that a
square cloud occults the stellar disk. In the second, we allowed the position
of the cloud to shift off the projected stellar disk so that it contributed
only line emission to the spectrum.
The dimensions of real clouds near a magnetic B star are actually likely
to be much larger than one stellar radius (Hunger et al. \cite{hunal1},
Bolton \cite{bol1}). Thus, {\it we expect that our treatment underestimates
the total effects of emission at non-occultation phases.}
However, since we modeled only
differences in line strengths between occultation and non-occultation phases,
this fact is not relevant to our analysis.

  The density and areal coverage factors of the cloud are necessary but not
critical parameters in our analysis. For our initial models we started with
a trial volume density of 10$^{12}$ cm$^{-3}$, based on the appearance of high
Balmer series members during the occultation phases of $\sigma$\,Ori\,E
(Groote \& Hunger \cite{grohun1}, Short \& Bolton \cite{shobol}). From the
resulting column
densities ($\S$\ref{specphot}) and an estimate of $6 R_{*}$ for the extent of
the cloud in the line of sight from these authors, we derived a mean density
of $\sim$3--5$\times10^{10}$ cm$^{-3}$. As a compromise
we chose a density of 1$\times10^{11}$ cm$^{-3}$,
which we used in our models. The results of this study are not very sensitive
to the electron density for the range considered.

 {\it CIRCUS} includes a provision for various doppler effects. The most
important of these is stellar rotation, which {\it CIRCUS} treats by
shifting the local intensity spectrum by a wavelength equivalent to the
projected doppler shift at each point on the disk.
The program computes the net doppler
velocity between each projected element of the cloud and the background star
along the observer's line of sight. In our model there are no differential
velocities between regions of the cloud along the line of sight and the
projected area on the star because the cloud is co-rotating. Additionally,
because microturbulence can be an additional factor in determining the
signatures of a cloud on a spectrum, we varied this as a free
parameter. In general, values less than 20 km\,s$^{-1}$ resulted in similar
cloud absorption/emission spectra, so we used this value as a default.

\section{Basic picture }
\label{sketch}

  To date only a few models of the dynamics of rotating magnetic B stars have
been discussed for He-variable stars. Of these most work has focused on
$\sigma$\,Ori\,E. Borrowing from work of Groote \& Hunger (\cite{grohun1},
\cite{grohun2}, \cite{grohun4}; Hunger \& Groote \cite{hungro1};
Hunger et al. \cite{hunal1}), Shore and Brown (\cite{shobro}), Babel \&
Montmerle (\cite{babmon1}, \cite{babmon2}), and Babel (\cite{bab}), this
prototype and probably
all He-variable stars exhibit variations because they have dipolar fields
which are inclined to some extent with respect to their magnetic axes. There
is now a broad consensus that closed magnetic loops from the two poles divert
a weak wind emerging from these surface regions.
The particles emanating from
opposite poles collide at high velocity and form a torus-shaped magnetosphere.
The particles eventually leak out through the outer edge of the torus
(defined by the Alfv\'en radius) and also settle back to the star -- to what
relative extents is still unclear.

The radiative wind theory of Springmann \& Pauldrach
(\cite{sprpau}) suggests that there is a tendency in this region of
$T_{\rm eff}$, log\,$g$ space for H and He atoms to separate during their flow.
This chemical fractionation is one of several complex feedback processes
which determines the evolution of the wind.
In our picture a
relative preponderance of metal ions are directly accelerated by the magnetic
polar wind along open field lines and leaves the star
permanently. At the same time many helium atoms are decoupled early in the
flow (as is hydrogen in cooler stars) and return back to the surface along
roughly the lines of force that guided their upward flow. At the surface
{\em Bp-like}
chemical anomalies (He-richness, metal-deficiency) establish local vertical
abundance gradients over time unless they are destroyed, e.g. through mixing.
Wind particles guided along closed lines accumulate in the torus for a short
time before they are lost at their edges.
The length of time these particles spend in the torus is far shorter
than the evolutionary lifetime of the star (e.g., Groote \& Hunger \cite{grohun2},
Havnes \& Goertz
\cite{havgoe}, Linsky et al. \cite{linal1}), so the torus's composition
will mirror the abundances found at the star's magnetic poles.

\begin{table}[th]
\begin{center}
\begin{tabular}{l|c|c|c|c}
\hline\hline
     & $\sigma$\,Ori\,E   &  HD\,184927  &  $\beta$ Cep & HR\,6684 \\
\hline
T$_{\rm eff}$ & 23\,000$^{1}$ & 23\,000$^{2}$ & 27\,000$^{3}$ & 26\,000 \\
log\,$g$       & 4.0        & 4.0        & 4.0        & 4.0        \\
$v$\,sin\,$i$  & 162$^{4}$        & 15$^{2}$         & 30 $^{5}$      & 100\\
$P_{rot}$  & 1.1908106$^{6}$  & 9.52961$^{2}$ & 12.00092 $^{7}$  &
$\approx$3.75$^{8}$ \\ $\Phi=0.0$    & 44241.537 & 49987.388 & 51238.15 &
49953.03 \\ \hline
\end{tabular}
\end{center}
\caption{ \label{tab1} Adopted Atmospheric Parameters. NOTES:
(1) $T_{\rm eff}$ is given in K, $v$\,sin\,$i$ in km\,s$^{-1}$, and the
zeropoint of rotational phase (North magnetic pole crossing) as HJD + 2400000.
(2) $\Phi=0.0$ for HR\,6684 might be quite inaccurate.
(3) Sources are given as indices:\newline
{1 Groote \& Hunger \cite{grohun4},
2 Wade et al. \cite{wadal1}, 3 Gies \& Lambert \cite{gielam},
4 Bolton et al. \cite{bolal1}, 5 Campos \& Smith \cite{camsmi},
6 Hesser et al. \cite{hesal1}, 7 Henrichs et al. \cite{henal2},
8 Henrichs et al. \cite{henal1}.} }
\end{table}

  The above description is meant only to lay out a broad picture that is
pertinent to this paper. Havnes \& Goertz (\cite{havgoe}) and
Linsky et al. (\cite{linal1})
have discussed dynamic models of a magnetic {\em plasmasphere} in which
nonthermal electrons emit gyro-synchrotron radiation in a current sheet
some 10--20 $R_{*}$ from the star (Phillips \& Lestrade \cite{philes}).
These models were conceived in part because of the abnormally high
radio fluxes of these stars (Linsky et al. \cite{linal1}). In contrast,
X-ray observations do not suggest an abnormal level of high energy
emission (Drake et al. \cite{draal2}).
This same comment applies to the strength of the wind
components of the ultraviolet resonance lines.
Moreover, at a given rotational phase the variability patterns of
the ultraviolet and optical lines
appear very similar from one epoch to another at any given rotational phase.
For these reasons the volumes emitting fluxes in these different wavelength
regimes are probably not co-spatial. Thus, the physics of this emission
is likely to be quite different.
In this paper our philosophy will be to infer kinetic
and thermodynamic properties of the magnetospheric torus-clouds within
several stellar radii using the ultraviolet and optical data alone.
Additionally, we will be reluctant to invoke magnetic (or rotational) energy
sources or sinks unless needed.

  Our examination of these questions proceeds by considering first
the obscurations of the photospheric spectrum by the torus-cloud.
Since the clouds are locked into co-rotation with the surface,
the photospheric and cloud contributions of a line profile cannot be
separated according to differences in their shapes. Instead we must rely
on the modulation of the cloud's absorptions when it occults the star
once or twice each rotational period. Before proceeding to this analysis,
we first discuss the relevant parameters required to compute the
underlying photospheric line spectrum.

  Ultraviolet studies have shown that the He-strong stars
are stars on or just evolving off the main sequence (e.g., Shore \cite{shor})
and hence have log\,$g$ values in the range 3.5--4.0.
We chose  $T_{\rm eff}$ values for our program stars from relevant papers in
the literature and list them in Table\,\ref{tab1}.
Rotational periods and magnetic ephemerides are well known for these
stars (except for HR\,6684) and are listed as well.

\section{Spectral analysis of line absorptions }
\label{analy}

\subsection{The photosphere: metallicity and T$_{\rm eff}$ }
\label{photmet}

  Hunger et al. (\cite{hunal1}) and Reiners et al. (\cite{rein})
concluded that the metal abundances on the surfaces of $\sigma$\,Ori\,E,
are low at the magnetic poles and roughly solar-like over the
magnetic equator. However, their analyses were performed on optical spectra
with the implicit
assumption that the effects of C\,II, Si\,II line absorptions from
the magnetospheric clouds are negligible. The Reiners et al.
metallic-line strength curves show the same offset in phase with respect
to the magnetic pole crossing phase that our Fig.\ref{starewlines}a does,
suggesting that enhanced optical and ultraviolet absorptions are formed
in the same locality. We tested this possibility by constructing additional
{\it CIRCUS} models for the optical C\,II and Si\,II wavelengths and found
that these were able to reproduce fractional equivalent width changes of
10--20\%.  Thus we are led to a contrary picture, namely that the
variation of UV and optical lines and the continuum is due to absorptions
within a co-rotating cloud.

\begin{figure*}[thp]
\psscalefirst
\centerline{\psfig{figure=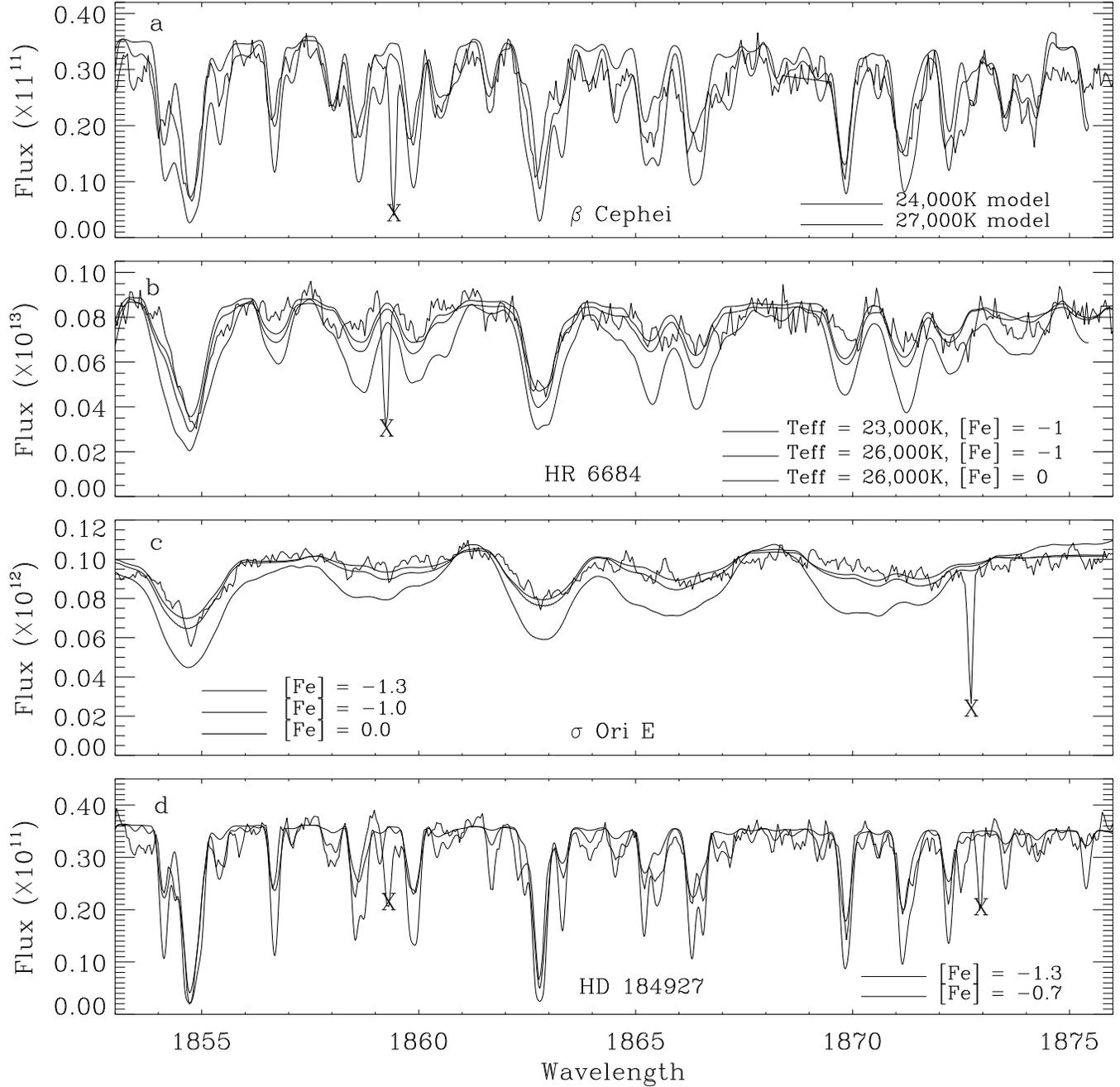,height=180mm,angle=90}}
\caption{{\bf a-d}
Comparison of {\it IUE} observations with {\it SYNSPEC}-synthesized
spectra of the wavelength region surrounding the Al\,III doublet.
The observations were made near the phase when the North magnetic pole
crossed the stellar meridian:
{\bf a}\, for $\beta$\,Cep, spectrum SWP 52432 is compared with SYNSPEC models
having $T_{\rm eff}$ = 27\,000\,K, [Fe] = 0, and $T_{\rm eff}$ = 24\,000\,K,
[Fe] = -1,
{\bf b}\, for HR\,6684 (at line minimum; no field detection yet) spectrum
SWP 55488 is plotted with models having $T_{\rm eff}$ = 26\,000\,K,
[Fe] = 0, -1, and $T_{\rm eff}$ = 23\,000\,K, [Fe] = -1,
{\bf c}\, for $\sigma$\,Ori\,E, five spectra (SWP 07555--6, 15785--7) are
compared with models having $T_{\rm eff}$ = 23\,000\,K, [Fe] = 0, = -1 \& -1.3,
{\bf d}\, for HD\,184927 four {\it IUE} spectra (SWP 14504, 14511, 55612,
55770) are compared with $T_{\rm eff}$ = 26\,000\,K, [Fe] = -0.7 \& -1.3;
the latter model fits so well it is largely hidden by the observed spectrum.
}
\label{staral3}
\end{figure*}

  Our line synthesis
analysis was conducted by first selecting rotational phases for which
line strengths in our IUE data were minimal and then using {\it SYNSPEC}
to match spectra in the 1853--1876 and 1906--1931 Angstrom regions. We chose
$\beta$\,Cep as a control star and discovered that the lines
can be nearly similarly matched by model atmospheres having either a
normal metal abundance and a high effective temperature (27\,000\,K) or a
metallicity of one-tenth
the solar value and a lower temperature (24\,000\,K). We were able to break
this ambiguity by modeling the wings of the Si\,III
$\lambda$1206 and Si\,IV $\lambda$1394, $\lambda$1403 lines only with
the higher temperature model. This discrimination is possible
because the ionization of Si$^{2+}$ and hence the strength of
$\lambda$1206 is more sensitive to $T_{\rm eff}$ than to metallicity.
The high-temperature, normal-abundance model leads to
less extended wings for $\lambda$1206. Our adopted
value $T_{\rm eff}$ = 27\,000\,K (and normal composition) agrees well
with the value 26\,600\,K found by Gies \& Lambert \cite{gielam}.
Fig.\,\ref{staral3}a shows overplots
of an {\it IUE} spectrum in the region of the Al\,III doublet with
both the adopted and the 24\,000\,K synthesized models.

  Following the same procedure, we estimated the metal abundances for a
minimum-absorption observation (SWP 55488) of HR\,6684 in the Al\,III and
Fe\,III wavelength regions. As with $\beta$\,Cep, we found
a trade-off between composition and stellar effective temperature, but
this time both a low abundance and high $T_{\rm eff}$ was needed to simulate
the comparatively weak line strengths.
As before, the weak wings of the Si\,III and Si\,IV resonance lines at
the non-occultation phases confirm this model.
The fitting ambiguity was again settled in favor of high temperature
(26\,000\,K) but a low metal abundance.
Fig.\,\ref{staral3}b  exhibits the
comparison for both low ([Fe]= -1) and normal metallicty and this effective
temperature as well as a low-metals, high temperature model. The metal lines
for the normal-abundance, low-temperature case are of course even stronger.
For both effective temperatures the normal abundance model can be ruled out.
For HR\,6684 we estimate a
a metallicity [Fe] = -1.0 $\pm{0.3}$ dex (internal errors).

  We digress
to point out that because helium-rich atmospheres have comparatively low
continuous opacities, a synthesized spectrum matched to observed metal
line strengths will tend to err on the {\em high}
side of the true photospheric metallicity. However, such errors
are likely to be small for the mild helium overabundances of 2--3$\times$
as occur on most He-strong stars. A similar assessment can be made of
the effects of magnetic fields to the measured line splitting, either due
to Zeeman splitting or to the modification of the atmospheric structure.
Such effects could produce in principle a line strengthening when the
magnetic poles are visible and thus would produce small-amplitude line
strengthenings at this phase. Note that such effects go in the wrong
direction to explain the general line-strengthening during the
occultation phase.

  Our fits to lines of the He-strong stars $\sigma$\,Ori\,E and HD 184927
were conducted by using the effective temperatures, 23\,000\,K, from
previous optical studies. For these stars the abundances refer explicitly
to the magnetic polar caps. In Fig.\,\ref{staral3}c
we depict the synthesized spectrum in the Al\,III line region with a
low-absorption observation of $\sigma$\,Ori\,E at $\phi$ $\approx$ 0.8,
that is, between the crossings of the North magnetic pole and the associated
He-spot.
For broad-lined spectra such as these, the accuracy of an abundance
determination deteriorates. Nonetheless, our comparisons show a preference
for the [Fe] = -1 $\pm{0.4}$, or less.
Thus, our results are in mild disagreement with the optical
result with the Hunger et al. (\cite{hunal1})
finding of a metal deficiency of only a factor of 2.5 but are comparable
to the Reiners et al. (\cite{rein}) results.  The program star, HD\,184927,
provides an optimal opportunity for line synthesis comparisons because
its photospheric lines are virtually unbroadened.
Fig.\,\ref{staral3}d exhibits a match with an average
of four IUE spectra in the Al\,III line region with a metal
abundance [Fe] $\approx$ -1.3$\pm{0.3}$.

  We now consider line formations in the clouds, starting first
with the weak metallic line absorptions formed at great distances from the
star and will then work our way inwards.

\subsection{Comparison of time-variable spectra }

  Another approach to disentangling the spectra of
the cloud and photosphere is by means of the Temporal Variance Spectrum
({\em TVS}). Its use opens the question of whether heterogeneous metal
abundances across the star's surface contribute appreciably to the observed
variation of metallic line strengths.
This spectrum permits a comparison of each line's {\em activity}
with its mean strength. If photospheric abundances change from point to
point on the star, the absorptions of all lines will be affected to some
degree. On the other hand,
if excess line absorptions occur in a cool circumstellar cloud at some
distance from the star, low-excitation (and usually strong) lines
are those we expect primarily to show fluctuations in the TVS.

\begin{figure*}[thp]
\psscalefirst
\centerline{\psfig{figure=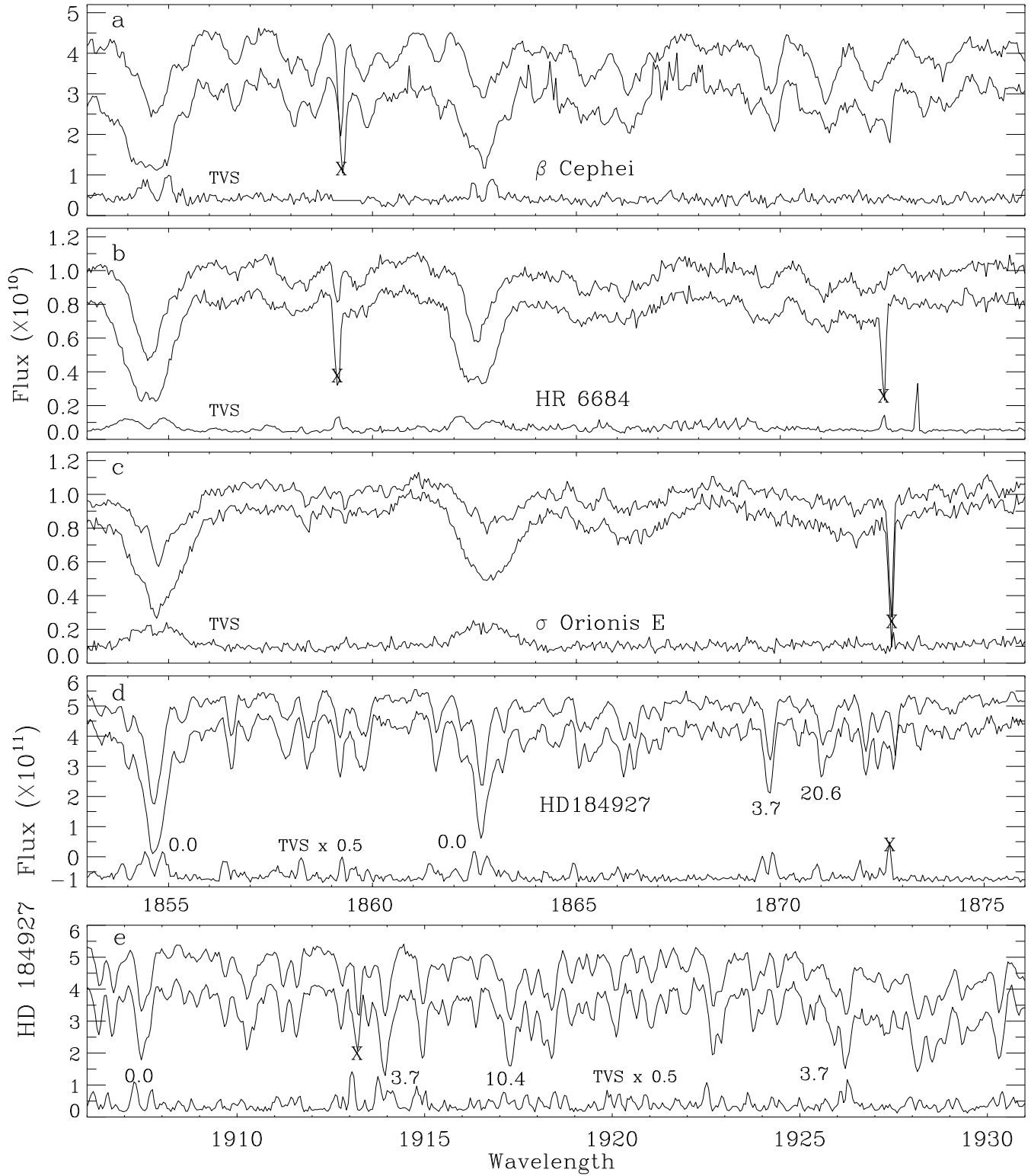,height=180mm,angle=90}}
\caption{{\bf a-e}
Comparison of spectra in Al\,III doublet region for program stars;  {\em clear
star} and occultation phases are shown in the top and lower spectrum
(spectra offset for clarity). The
TVS spectrum is shown at the bottom for the ensemble of spectra:
{\bf a}\, for $\beta$\,Cep, upper spectrum is from  SWP 06234, 42939, \& 52620,
and the lower one is from SWP 46244, 46255, 52121, 52460, 52573, \& 52594,
resp.,
{\bf b}\, for HR\,6684, upper spectrum is from SWP 14514, 55668, 55672, \&
55732,
{\bf c}\, for $\sigma$\,Ori\,E upper spectrum is from SWP 07553, 07555,
and the lower one from SWP 07534, 07536--7, 07560, 07583, \& 07609,
{\bf d}\, for HD\,184927 the upper spectrum (top) is from SWP 14468, 14475,
14487, 55669, 55685, 55701, and the lower one is from  SWP 14504, 14511, 14681,
55629, 55770, \& 55781 (excitations are given for a few lines),
{\bf e}\, for HD\,184927 as in {\bf d} but for the wavelength range
surrounding low-excitation Fe\,III lines.
}
\label{starmetstvs}
\end{figure*}

  Figs.\,\ref{starmetstvs}d,e test this expectation for HD\,184927.
Because of its slow rotation this star offers the best case
for detecting the variability of individual lines.
In this figure we display two spectral orders containing prominent
Al\,III and Fe\,III lines. Because the excitation potentials of
the lower levels of several other Fe\,III (and Ti\,II) lines are high,
we give these values as well.
The lower (TVS) plot shows that only those lines with excitations of a
few eV or less exhibit significant TVS fluctuations.
Fig.\,\ref{starmetstvs}c shows the same pattern
for the Al\,III line region of $\sigma$\,Ori\,E. Clearly, all or nearly all
the variations of the metal lines arise in a cool medium. Inspection of the
line depths confirms that they arise one-quarter of a cycle earlier or later
than the transits of the star's meridian. 

To place limits on the
variation of metals on the surface of HD\,184927, we chose three IUE spectra
obtained at faint-star (magnetic null) phases and compared line synthesis
models containing weak, high excitation lines. Our best fitting models for the
spectra in the two orders containing the Al\,III and Fe\,III lines
suggested a metallicity
of [Fe] = -1.0$\pm{0.3}$ (internal error), which is only marginally
different from the metallicity of -1.3 estimated from
Fig.\,\ref{staral3}d.
The reader can discern the line activity from the lower line in
Fig.\,\ref{starmetstvs}d, which is the TVS spectrum of all observations.
Note that the Al\,III lines themselves still
have a residual absorption, which we interpret as due to the cloud.
Otherwise, the
differences for most other (high-$\chi$) lines are generally nil. Plots for
the spectrum in the Fe\,III $\lambda$1914 region show similarly small
differences. We conclude that essentially all the UV line variations
are due to absorptions from occulting clouds. Supporting evidence for
this conclusion comes in addition from near-UV and visible-band data for
$\sigma$\,Ori\,E discussed in the next section.

   As Fig.\,\ref{starmetstvs} shows, the TVS profiles have double lobes.
To investigate the cause of these structures, we averaged two subgroups
of spectra observed at phases of maximum and minimum strengths (hereafter
referred to as {\em clear-star} and {\em occultation} phases).
The low-excitation
lines in the clear-star spectra have triangular shapes and narrow cores in
these plots whereas the occultation-phase spectra exhibit broad, rounded cores.
We confirmed that these
differences in shapes arise during particular phases by inspecting individual
spectra and seeing the patterns in them.
Because the cloud-tori are the presumed products of a rapid deceleration
of wind particles, it is natural to attribute the excess line broadening
as a part of the deceleration process - a turbulence of 25--50 km\,s$^{-1}$.
\footnote{It is conceivable that this broadening could instead have a
magnetohydrodynamic origin, e.g. from the generation of Alfv\'en waves by
motions of field lines. However, for a dipolar field in a weak-field star
like $\beta$\,Cep the velocities of such waves would probably be smaller
than these values. Moreover, Alfv\'en waves would produce motions transverse
to the line of sight when the torus occults the star.}

\subsection{The cloud spectrum and derived cloud parameters }
\label{specphot}

\begin{figure}[tp]
\psscalefirst
\centerline{\psfig{figure=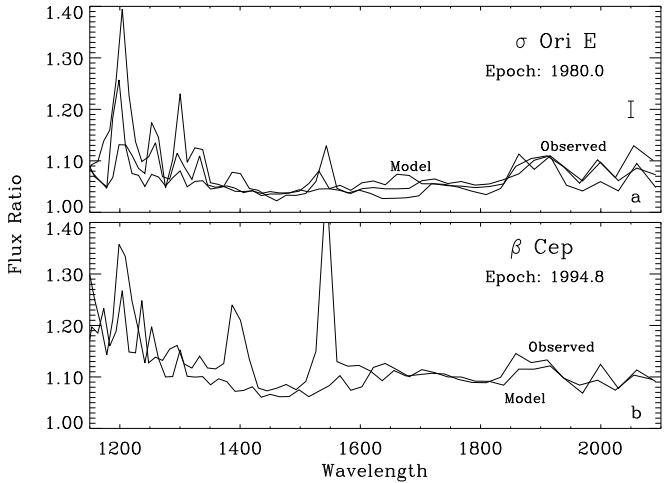,height=88mm,angle=90}}
\caption{{\bf a and b:}
Observed and computed ratios of (quantity + 1) cloud absorption spectra for
$\sigma$\,Ori\,E and $\beta$\,Cep. {\bf a} for $\sigma$\,Ori\,E
the observed curve (thick solid line) is the ratio of {\it IUE} spectra
SWP 07555-6, 07558, 07560 ($\phi$ = 0.40--0.59) to SWP 07583 ($\phi$=0.02,
observed several hours earlier), binned one point per echelle order.
Dot-dashed and dashed line curves denote {\it CIRCUS} models with T$_{cloud}$
= 11\,500\,K and 13\,000\,K, with column densities of 6$\times$10$^{22}$
and 1$\times$10$^{23}$, and with coverage factors of 100\% and 50\%,
resp. The cloud spectra represent the fractional
absorption plus one (in continuum units).
{\bf b} For $\beta$\,Cep the observations are taken from the ratio of
SWP 52415, 52594 to SWP 52514, 52653. The model has parameters
T$_{cloud}$ = 11\,500\,K, a column density of 1$\times$10$^{23}$
particles cm$^{-2}$, and a 70\% coverage factor.
}
\label{fig3}
\end{figure}

 {\it CIRCUS} was used to compute the absorption of the
stellar flux by the cloud (the so-called {\em iron curtain}) by
subtracting the photospheric spectrum, computed by {\it SYNSPEC} from the
photospheric plus cloud spectrum computed by {\it CIRCUS}. This was done
for a variety of trial temperatures, column densities, and occasionally
other parameters such as volume density, microturbulence and areal coverage
factor.
In general, the lower the cloud temperature, the
higher the monochromatic absorption because of the increased influence
of hydrogen opacity.
Fig.\,\ref{fig3}a shows a comparison
of the ratio of {\it IUE} spectra obtained at the maximum and minimum flux
phases for $\sigma$\,Ori\,E. The principal features of interest in this figure
are the increased absorption below 1300 Angstroms and also
near $\lambda$1900. Our tests showed that these wavelength-dependent
features can be reproduced by having cool temperatures, {\it viz.}
$T_{\rm cloud}$ = 11\,500 -- 13\,000\,K.
In Fig.\,\ref{fig3}b we show a fit with similar parameters
to observations of $\beta$\,Cep taken from
clear-star and occultation phases of the rotation cycle (but at similar
pulsation phases).
For both stars these cloud temperatures are
consistent with radiative equilibrium temperatures expected for material
situated several stellar radii from the surface (e.g. Drew \cite{dre},
Millar \& Marlborough \cite{milmar}). We also experimented with two-zone
cloud models.
Because inner-cloud temperatures are likely to be higher in these models,
and the opacities lower, the total cloud opacities at a given wavelength tend
to be much as the same as for one-zone, low-temperature models. As a result,
we found that for the weak Fe-curtain one cannot distinguish very well
between judiciously chosen multiple temperature and single low-temperature
models.

As an additional test of the models for Fig.\,\ref{fig3}a, we reran
the models for the $\lambda\lambda$3700--4000 range used by Groote \& Hunger
(\cite{grohun1})
to show that the torus occultation is accompanied by increased
core absorption of the high-level Balmer-lines.
Our models (not shown) exhibit similar line strengthenings
as those observed (20--30\%). This result further validates the 
interpretation that the line strength variations arise
from absorptions in the cloud.

\begin{figure*}[thp]
\psscalefirst
\centerline{\psfig{figure=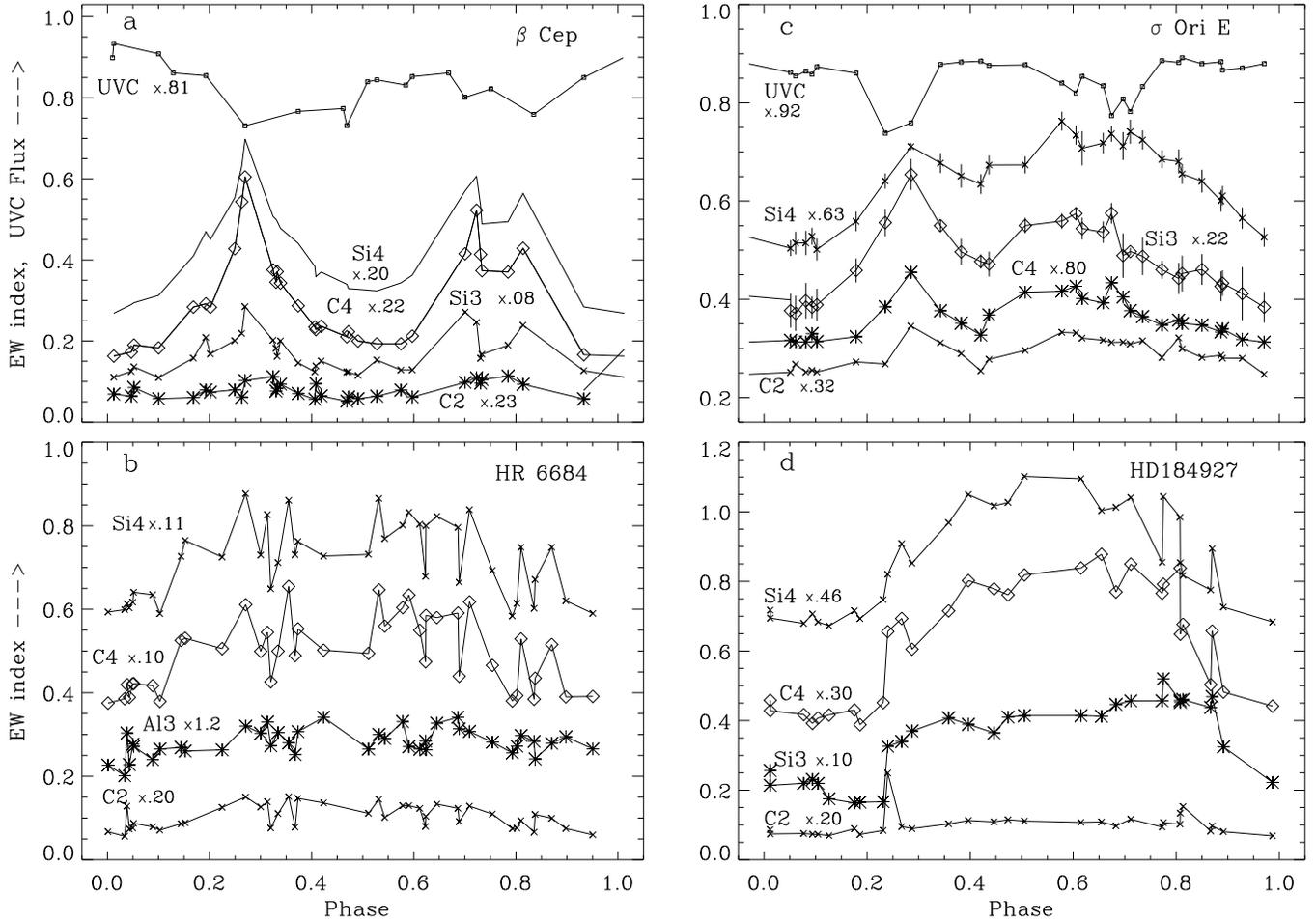,height=180mm,angle=90}}
\caption{{\bf a-d}
Equivalent width indices of various lines of the program stars with
rotational phase (periods given in text). In each case the index is defined
as the reciprocal of the {\it ratio} of the fluxes within a $\Delta\lambda$
bandpass around line center (Table\,\ref{tab2}) to the ratio of valid fluxes outside
these limits. The curves given are for Si\,IV $\lambda$1394, 1403 (mean),
C\,IV $\lambda$1548, Si\,III $\lambda$1206, C\,II $\lambda$1298.
The curves are renormalized by the factors shown in the
annotations and separated by vertical shifts.
}
\label{starewlines}
\end{figure*}

  To place our spectroscopic results in a context in which they can be
compared with previous photometric studies, we summed
the fluxes in four IUE echelle orders in the uninterrupted wavelength range
$\lambda\lambda$1800--1905 and plotted them as a function of time. The
resulting continuum ({\em UVC}) light curves for $\sigma$\,Ori\,E and
$\beta$\,Cep in
Figs.\,\ref{starewlines}a,c (top curves) exhibit
a pair of light dips corresponding to cloud occultations. The stronger
of the two dips is 14\% for $\sigma$\,Ori\,E and 11.5\% for
$\beta$\,Cep.\footnote{The
narrowness in phase of the dips for $\sigma$\,Ori\,E may eventually turn out
to be important in constraining the thickness of the torus in the direction
of the magnetic axis, at least at several stellar radii from the star.}
We estimate the corresponding flux dips for HR\,6684 and HD\,184927 to be
2$\pm{0.7}$\% and 3--4$\pm{1}$\%, respectively, i.e., just barely detectable.
Some interesting conclusions can be drawn about the chemical composition of
the $\sigma$\,Ori\,E cloud by comparing the amplitude of the primary dip in
this star's UVC light curve, 14\%, with the $\approx $11\% variation in the
Str\"omgren {\em u}-band (Hesser et al. \cite{hesal1}). To see what cloud
conditions could make the {\it IUE}- and
ground-based absorptions of $\sigma$\,Ori\,E consistent with one another,
we computed {\it CIRCUS} models in the $\lambda\lambda$1800--1905 and
$\lambda\lambda$\,3450-3650 (roughly, Str\"omgren-{\em u}) regions.
Several cloud parameters were varied, including temperature,
metallicity, turbulence, and column density. All our trials predicted
larger absorptions for $\lambda\lambda$1800--1905 region than for
the Str\"omgren-{\em u} region. The most important result of
these tests was that the absorption ratio of these wavelengths was found
to be highly sensitive to metallicity. For example, the ratio decreases
markedly from 2.0 at normal metallicity to 1.4 for [Fe] = -1.
This sensitivity can be explained as follows:
the fluxes of a B star in the {\em u}-bandpass are almost
independent of metallicity because they are determined almost entirely
by Balmer bound-free opacity. In contrast, in low-metal models the
contributions of hydrogenic and metal line opacities are roughly equal
in the $\lambda\lambda$1800--1900 region. Since the great majority of the
metal lines in this wavelength region are unsaturated, their absorptions vary
almost linearly with metallicity. Taking our computed ratio of 1.4 for
the low-metals case, these results suggest that observed {\it u}-band variations
should be about 1.4$\times$0.11 = 14.5\%, i.e.\, almost the same as
the observed value of 14\%.
In contrast, a model with a normal
cloud metallicity predicts a flux dip of about 22\% for
$\lambda\lambda$1800--1900,
which may be excluded easily by the observations. Altogether,
the metallic abundance of the cloud in $\sigma$\,Ori\,E
derived from a comparison of flux amplitudes in the middle-
and near-UV is in very good agreement with our line synthesis
analysis ($\S$\ref{photmet}). It is also in agreement with our general
picture which
requires that the chemical composition of the cloud should be essentially
identical with the source material from the outflowing regions of the
photosphere.

\subsection{Cloud parameters from weak resonance lines}
\label{weakres}

  In this and the next subsection we consider the formation of the
added absorptions of weak and/or strong resonance lines of silicon, carbon,
and nitrogen by using {\it CIRCUS} models to match the added absorptions
observed at occultation phases.
We will also discuss the necessary roles of turbulence and non-LTE
processes for strong resonance lines formed in the cloud.

\begin{table*}[t]
\begin{center}
\begin{tabular}{c|ccc|ccc|ccc|ccc}
\hline\hline
   &   & $\sigma$\,Ori\,E  &  & &  HD\,184927  &  & &  $\beta$ Cep & & &
HR\,6684 & \\
\hline
Ion [eV] &  EW & EW & $\Delta\lambda$ & EW & EW &
$\Delta\lambda$ & EW & EW & $\Delta\lambda$ & EW & EW & $\Delta\lambda$ \\
Line & Max  & Min  & (\AA)   & Max  & Min  & (\AA)   & Max  & Min  & (\AA)
& Max  & Min & (\AA) \\ \hline
C\,II [24.4] &  &   &   &   &   &   &   &   &   &   &  &  \\

1336 & .682: & .557: (.125) & 1.3  & .520: & .370: (.150) & 0.5 & .292:
& .245: (.047) & 0.6 & 1.00: & .880: (.120) & 1.1   \\
\hline
C\,IV [64.5] &  &   &   &   &   &   &   &   &   &   &  &  \\
1548 & .944 & .545 (--)
& 1.5  & .838 & .158 (--)
& 1.3 & .890 & .483 (--)
& 1.1 & 1.085 & .234 (--)
& 1.2   \\
1550 & .794 & .448 (--)
& 1.5  & .838 & .158 (--)
& 1.3 & .890 & .483 (--)
& 1.1 & .863 & .157 (--)
& 1.2   \\
\hline
N\,V [97.9] &  &   &   &   &   &   &   &   &   &   &  &  \\
1238 & .192 & .042 (--) & 1.5  & .421 & .158 (.530)$^{*}$ & 1.3 & .506
& -.097 (.750)$^{*}$ & 1.1 & .303 & -.035 (--) & 1.2   \\
\hline
Si\,II [16.3] &  &   &   &   &   &   &   &   &   &   &  &  \\
1265 & .198 & .150 (.048) & 1.0 & .385 & .372 (.013) & 1.0 & .245
& .248 (.00) & 1.1 & .171 & .164 (.007) & 1.2   \\
1309 & .122 & .074 (.048) & 1.0 & .356 & .316 (.040) & 1.3 & .090
& .071 (.019) & 1.1 & .288 & .268 (.020) & 1.2   \\
\hline
Si\,III [33.5] &  &   &   &   &   &   &   &   &   &   &  &  \\
1206 & .614 & .395 (.219) & 1.4 & .990 & .739 (.251) & 1.0 & .778
& .435 (.343) & 1.2 & 1.661 & 1.315 (.346) & 1.2   \\
1301 & .247 & .191 (.056) & 1.5 & .416 & .368 (.048) & 0.6 & .346
& .342 (.004) & 0.6 & .417 & .397 (.020) & 0.8   \\
1303 & .227 & .167 (.060) & 1.5 & .435 & .399 (.036) & 0.6 & .316
& .311 (.005) & 0.6 & .402 & .392 (.010) & 0.8   \\
1892 & .459 & .339 (.120) & 1.5 & .479 & .416 (.063) & 0.6 & .524
& .467 (.057) & 0.6 & .406 & .394 (.012) & 0.8   \\
1895 & .353 & .198 (.155) & 1.5 & .476 & .424 (.052) & 0.6 & .460
& .407 (.053) & 0.6 & .376 & .364 (.012) & 0.8   \\
\hline
Si\,IV [45.1] &  &   &    &   &   &   &   &   &   &   &   &  \\
1394 & .810 & .589 (.221) & 1.4 & .626 & .377 (.249) & 1.2 & .778 & .542 (.236)
 & 1.0 & 1.207 & .817 (.390) & 1.4 \\
1403 & .662 & .426 (.236) & 1.4 & .471 & .258 (.213) & 1.2 & .770 & .523 (.247)
 & 1.0 &  .871 & .711 (.160) & 1.4 \\
\hline

Al\,III [28.4] &  &   &   &   &   &   &   &   &   &   &  &  \\
1855 & .551 & .322 (.229) & 1.6  & .633  & .319 (.314) & 0.8 & .772
& .449 (.323) & 1.3 & .629 & .353 (.276) & 1.4   \\
1862 & .480 & .260 (.120) & 1.6  & .434  & .302 (.132) & 0.8 & .434
& .320 (.114) & 1.3 & .556 & .270 (.286) & 1.4   \\
\hline
\end{tabular}
\end{center}
\caption{ \label{tab2} Observed Equivalent Width Extrema,
NOTES: (1) Units of equivalent-width (EWs) and wavelengths
are Angstroms were measured for $\Delta\lambda$ range noted.
(2) Parenthesized entries after {\em Min}
are differences between maximum and
minimum EWs at midpoints of clear-star and occultation phases; a colon
denotes an uncertain value.
(3) Bracketed entries following the listed ion is the ionization potential
in eV.
(4) Blank entries for Si\,IV and N\,V indicate that models are not computed
for these ions in Table\,\ref{tab3}.
(5) N\,V entries (starred) for HD\,184927
and $\beta$\,Cep refer to full-profile absorptions and are modeled in
Table\,\ref{tab3}.
}
\end{table*}

  Table\,\ref{tab2} gives the extreme equivalent widths in Angstroms observed at
clear-star and occultation phases.
These strengths are measured with wavelength limits specified around line
center (generally roughly given by $\pm{{\it vsin\,i}}$)
in order to isolate spectroscopic variations of a cloud
from other circumstellar activity, such as the wind. The
technique of measuring line strength variations in specific wavelength bands
of {\it IUE} spectra accurately was
introduced by Shore \& Adelman (\cite{shoad}) for the He-strong
stars.
To gain some
perspective of how representative the variations are in the program stars,
we also measured the equivalent widths of the same lines in the archival
datasets of other He-strong stars studied by SB90 with known periods.
We found that the range of variations for $\sigma$\,Ori\,E is typical for
this subclass. For example, the range for the Si\,IV
doublet of $\sigma$\,Ori\,E is similar to that found for HD\,37017,
larger than for HD\,37776, and smaller than for HD\,64760.

\begin{table*}[t]
\begin{center}
\begin{tabular}{c|cc|ccc||cc|ccc}
\hline\hline

 & HD\,184927 & & & EW & & $\beta$ Cep & & & EW & \\

\hline
Ion/Line& $T_{\rm cloud}$ & $v_{turb}$ & CD1 & CD2  & CD3    & $T_{\rm cloud}$ & $
v_{turb} $  & CD1 & CD2 & CD3  \\
  C\,II  &                 &            & 1   & 3.2  & 10     &               &
            & 1   & 3.2 & 10    \\
\hline
1336\AA & 15\,000 &   20  &            &      .118  & {\it .146} & 17\,000   &
20 &            &      .119  & {\it .146} \\
        & 14\,000 &   20  &      .121  & {\it .151} &      .178  & 16\,000   &
20 &      .124  & {\it .150} &      .173  \\
        & 13\,000 &   20  & {\it .147} &      .169  &            & 15\,000   &
20 & {\it .152} &      .182  & {\it .199} \\
\hline
 N\,V    &              &            & .32 &  1   & 3.2    &               &
       & .32 &  1  & 3.2   \\
\hline
1238\AA & 31\,000 &   50  &      .252- &      .478- &     +.589  & 29\,000   &
50 &            & {\it .742} &            \\
        & 30\,000 &   50  &            &      .282  & {\it .511} & 28\,000   &
50 &            &      .585  &            \\
        &         &       &            &            &            & 28\,000   &
100 &            & {\it .742} &            \\
\hline
 Si\,II  &              &            & 1   & 3.2  & 10     &               &
       &  1  & 3.2 & 10    \\
\hline
1265\AA & 16\,000 &   20  &            &      .006- &     +.017  &  not      &
   &            &            &            \\
        & 15\,000 &   20  &      .005- &     +.020  &      .048  & modeled &
   &            &            &            \\
        & 14\,000 &   20  & {\it .017} &            &            &           &
   &            &            &            \\
        &         &       &            &            &            &           &
   &            &            &            \\
1303\AA & 15\,000 &   20  &            &      .026- &     +.068  & 18\,000   &
20 &            &            & {\it .020} \\
        & 14\,000 &   20  &      .024- &     +.068  &            & 17\,000   &
20 &            &      .009- &     +.030  \\
        & 13\,000 &   20  &      .062  &            &            & 16\,000   &
20 &      .012- &     +.035  &      .075  \\
\hline
 Si\,III &             &             & 1   & 3.2 & 10     &               &
      &  1  & 3.2 & 10    \\
\hline
1206\AA & 16\,000 &    50 &            &      .216- &     +.260  & 17\,000   &
20 &            &            &      .259  \\
        &         &       &            &            &            & 17\,000   &
35 &            & {\it .330} &            \\
        &         &       &            &            &            & 16\,000   &
20 &      .248  & {\it .348} &            \\
        &         &       &            &            &            & 16\,000   &
35 & {\it .345} &            &            \\
        &         &       &            &            &            & 15\,000   &
20 &            &      .248- &     +.368  \\
        &         &       &            &            &            & 15\,000   &
30 & {\it .334} &      .492  &            \\
        &         &       &            &            &            &           &
   &            &            &            \\
1301\AA &17\,000  &    20 &      .029- &      +.048 &            & 19\,000   &
20 &            &            &      .006  \\
        &16\,000  &    20 & {\it .042} &            &            & 18\,000   &
20 &            &      .003- &     +.009  \\
        &15\,000  &    20 & {\it .044} &            &            & 17\,000   &
20 & {\it .004} &            &            \\
        &         &       &            &            &            & 16\,000   &
20 &      .007  &     (.03)  &     (.12)  \\
        &         &       &            &            &            &           &
   &            &            &            \\
1892\AA &16\,000  &    20 &            &      .032- &     +.084  & 20\,000   &
20 &            &            &      .052  \\
        &15\,000  &    20 &            &      .040- &     +.097  & 19\,000   &
20 &            &      .041- &     +.096  \\
        &14\,000  &    20 &            &      .038- &     +.092  & 18\,000   &
20 &      .032- &     +.079  &            \\
        &         &       &            &            &            & 17\,000   &
20 & {\it .063} &            &            \\
\hline
Al\,III  &             &             & 1   & 3.2 & 10    &                &
      &  1   & 3.2 & 10    \\
\hline
1855\AA & 18\,000 &   50  &      .177  &      .260- &     +.382  & 17\,000   &
50 &      .278  & {\it .330} &  \\
        & 17\,000 &   20  &            &            &      .246  & 16\,000   &
20 &            &      .258  & {\it .321} \\
        & 17\,000 &   50  &      .253- &     +.432  &            & 16\,000   &
50 &      .315- &     +.340  &            \\
        & 16\,000 &   20  &            &   .241     & {\it .312} & 15\,000   &
20 &            &      .281  & {\it .319} \\
        & 16\,000 &   50  &      .386  &            &            & 15\,000   &
50 &      .301- &     +.349  &            \\
\hline
\end{tabular}

\caption{ \label{tab3} Computed Cloud Absorptions. NOTES:
(1) Cloud temperatures are given in Kelvins, and microturbulences in
km\,s$^{-1}$.
(2) Columns ({\em CD1--3}) denote column density (units of 10$^{22}$\,cm$^{-2}$)
for models.
(3) Note that the absorption entries are computed from
the same $\Delta\lambda$ limits (except for N\,V) as given in Table\,\ref{tab2}.
(4) Pairs of entries close to observed values in Table\,\ref{tab2} are
italicized.
(5) Computed EWs are often somewhat below or above the observed value. In
such cases the table entries have
trailing (-) or preceding (+) symbols, respectively.
All models have coverage factors of 100\%.}
\end{center}
\end{table*}

  For the weak resonance lines
the differences between the observed equivalent widths in Table\,\ref{tab2}, in
parentheses, is
taken to be the total absorption contribution in the star's cloud(s).
This amount can be compared directly
with the detailed absorption predictions of {\it CIRCUS},
given in Table\,\ref{tab3} for $\beta$\,Cep and HD\,184927.
For a stated variation of a given line, the star-to-star differences can arise
from a variety of factors, but the dominant one is the star's metallicity.
Thus, similar cloud absorptions for lines of the similarly
metal-deficient stars HD\,184927 and HR\,6684 will lead to similar derived
cloud parameters. We also note that
the dependence of the derived cloud parameters on
metallicity is especially important for the strong resonance lines for which
the greatest contrast exists between a (weak) photospheric line and a
highly saturated cloud line.

\subsubsection{C\,II, Si\,II and weak Si\,III lines}
\label{siabs}

In Table\,\ref{tab4} we summarize our findings from Tables\,\ref{tab2}
and \ref{tab3} by giving
temperatures derived from C\,II, Si\,II, weak Si\,III, and N\,V
absorptions. We kept microturbulence constant (20 km\,s$^{-1}$,
except for N\,V) in this compilation and took entries from the middle of
the limits given in Table\,\ref{tab3}.
Before consolidating these results, we
discuss the status of the suitability of these lines in terms of possible
practical problems.

  The C\,II $\lambda\lambda$1335--6 doublet is a well-known temperature
diagnostic for cloud-like conditions, but in {\it IUE} spectra the 1335\AA\,
line is marred by the presence of an instrumental {\em reseau}.
In the spectra of $\sigma$\,Ori\,E the broadened wings of the 1336A
feature are blended by nearby low excitation Cr\,II and Fe\,II
photospheric lines, rendering $\lambda$1336 unusable for this star.
Both lines are further marred by an interstellar component (particularly for
$\sigma$\,Ori\,E). Despite these problems the C\,II lines noticeably broaden
in occultation phase spectra of all our stars. Reference to Table\,\ref{tab2}
shows that the $\lambda$1336 variation is moderately
large for HD\,184927  and HR\,6684. Indeed, these are the only C\,II results
that can be quoted with reasonable reliability.

  {\it IUE}/SWP spectra also include the Si\,II
resonance lines at 1264.7\,\AA, 1265.0\,\AA, and 1309.3\,\AA. The observed
variations for the doublet are immeasurably small for all the program stars
except $\sigma$\,Ori\,E. For $\beta$\,Cep, HD\,184927, and HR\,6684 the small
variations mean that the Si$^{1+}$ population is low because the cloud
temperature is high (or, less likely, very low)

  Variations in both weak resonance and low-excitation Si\,III lines
provide convenient cloud temperature diagnostics.
Consider first the forbidden Si\,III $\lambda$1892 resonance line, for
which the log\,$gf$ is 30\,000 times weaker than the $\lambda$1206 line's.
This line is weakly visible in spectra of normal early B-type stars
(Walborn et al. \cite{walal2}), but it strengthens for our
program stars during occultation phases. This is not surprising:
in contrast to the photosphere, one expects a cool circumstellar gas with 
a long path length and a shallow temperature gradient to produce a strong
absorption in a resonance line.

  The Si\,III lines also shed light on the cloud density.
Consider that the $\lambda$1892 resonance line in turn feeds the excited
lower levels of the $\lambda\lambda$1294--1303 multiplet of Si\,III. 
Both the observations (Table\,\ref{tab2}) and the {\it CIRCUS} models
demonstrate that the variations of the Si\,III resonance and multiplet 
features are comparable and well correlated among themselves.
These facts suggest that metastability of the $^{3}$P$^{o}$3p (lower) level
does not determine its
atomic population and thus that the cloud densities are moderately high
($\ge$ 10$^{11}$ cm$^{-3}$).

   To summarize this section, the ionization potentials of Si$^{1+}$,
C\,$^{1+}$, and Si\,$^{2+}$ (Si\,II, C\,II, Si\,III: $\chi_{IP}$ = 16--34 eV)
together provide an excellent bridge of diagnostics for conditions between
the inner and outer cloud regions. From Table\,\ref{tab4} one can see that the
line-formation temperatures decrease with ionization potential of the
parent ion, as one would
expect for a circumstellar cloud having an outward-decreasing temperature.

\subsection{ Resonance line absorptions as diagnostics}
\label{resabs}

\subsubsection{Probing temperature and geometry }

   According to
our {\it CIRCUS} models, the ionization of silicon and carbon most favors line
formations at temperatures centered at 21\,000\,K and 26\,000\,K, respectively,
for the expected cloud densities. These temperatures are comparable
and slightly higher, respectively, than what one might expect from
radiative equilibrium models for plasma near stars having $T_{\rm eff}$'s of
23\,000--27\,000\,K. Thus although the resonance lines are not
necessarily {\em superionized} features, they must be formed
in a warm environment, thus close to the star. As noted by
various authors (Fischel \& Sparks \cite{fisspa2}, 
Barker et al. \cite{baral1}, Shore \& Brown \cite{shobro},
Henrichs et al. \cite{henal1}), the strong resonance lines of C\,IV, Si\,IV and
other ions undergo dramatic changes during the rotational cycle of each of our
program stars.
Fig. \ref{starewlines} depicts the phase variations of an equivalent width
{\it index} for several strong resonance lines for each program star.\footnote{
The depicted UVC curves for $\beta$\,Cep and $\sigma$\,Ori\,E are time 
sequences of binned fluxes constructed from the echelle orders extending over
$\lambda\lambda$1800--1905; errors are estimated to be $\pm{3}$\%.
The $\beta$\,Cep UVC data are large-aperture observations of 1991-4 and are
corrected for a 5\% sinusoid found in the raw data at the pulsation period.
The $\sigma$\,Ori\,E UVC fluxes are from 1980-1.}
In each case this index is the ratio of the measured fluxes outside and inside
the $\Delta\lambda$ wavelength limits for that line, as specified in Table\,\ref{tab2}.
Similar indices computed from the blue and red halves of these
wavelength windows give in all cases essentially the same fractional amplitudes.
These checks indicate that the influences of the wind at moderate velocities
(or in the case of the C\,IV lines, the Fe\,III interstellar components) do
not contribute appreciably to the variation of this index.
The sequence of these strong line variations may also be compared
qualitatively to the variations of the Al\,III resonance doublet by consulting
Fig.\,\ref{staral3}.

\begin{table}[tb]
\begin{center}
\begin{tabular}{c|c|c|c}
\hline\hline
        & column  & HD\,184927 & $\beta$ Cep \\
Line    & density &            &             \\
\hline
C\,II   & $1\cdot 10^{23}$ & 15\,000\,K &  (none)     \\
Si\,II  & $3\cdot 10^{22}$ & 15\,000\,K & 16\,500\,K  \\
Si\,III & $3\cdot 10^{22}$ & 16\,500\,K & 18\,500\,K  \\
N\,V    & $1\cdot 10^{22}$ & 30\,500\,K & 28\,500\,K  \\
\hline
\end{tabular}
\end{center}
\caption{ \label{tab4} Synopsis of Best-Fit Cloud Temperatures for
Lines of Ions Listed (from Table\,\ref{tab3}).  NOTES:
(1)Typical column densities are in middle of best-fit parameter space.
(2) Errors are typically +/-1000K except for C II(ISM and reseau flaws)
and Si III ($\beta$\,Cep).
(3) Typical microturbulence values are 20 km\,s$^{-1}$, except for
N\,V (50 km\,s$^{-1}$).}
\end{table}

 The data in Fig.\,\ref{starewlines} are well enough sampled to draw
several conclusions concerning the geometry of the tori.
First, the absorptions among different strong resonance lines
peak at the same small ranges of phase.
This fact suggests that these lines form over the same stellar longitude.
However, for $\beta$\,Cep and $\sigma$\,Ori\,E
the ultraviolet quasi-continuum curves
are offset from the central absorption
phases of the strong resonance lines. This fact implies
a {\em warping} at large distances from the star
(see also Groote \& Hunger \cite{grohun4}, GH97).
Another conclusion from the sharpness
of the maxima in these figures is that the heights of the clouds
(perpendicular to their plane of symmetry) are at least one
 stellar diameter.
Otherwise, for short heights, their optically thick absorptions
would be proportional to their projected areas against the
stellar disk. Such a geometry would impose an (unobserved) local minimum
on the resonance-line absorption curve at the phase of mid-occultation.
We used this fact to adopt an area
coverage factor of 100\% in our {\it CIRCUS} analysis of cloud parameters.

\subsubsection{Probing the wind}

  To investigate the behavior of the blue-wing absorption from winds in
these stars, we modified our equivalent width extraction routine
to compute flux ratios
in a 2\AA\, segment of this wing to all valid fluxes in the order {\it
except those from the photospheric line profile}. With the contribution of
the latter suppressed, we found no variation (i.e., to $<$10\%)
in the wings of either the Si\,IV or C\,IV lines with rotation phase among any
of the program stars. Additionally,
the blue edge velocities are stable over the cycle
and are also remarkably uniform from star to star. GH97 had already found that
the {\em coronal} wind in $\sigma$\,Ori\,E is phase-independent, and the
constancy we find tends to extend this conclusion to our other stars. This is
a notable point because previous investigators have stated that the wind
fluxes of HD\,184927 vary from this star's magnetic pole to equator (Vauclair
et al. \cite{vaual1}). This erroneous perception has been an important
supporting tenet for the model of {\it photospheric} diffusive settling of
helium (Michaud et al. \cite{mical1}). In addition, these observations
suggest that the initial wind column, dubbed the {\em jet} by SB90, is not
visible in these data, though indeed it must ultimately be visible at some
level of photometric precision because of the optical thinness in the far
blue wings of the photospheric lines. The difficulty of observing the jets
could mean that their velocities are spread out over a large range.

\subsubsection{Probing density and turbulence}

\begin{figure*}[thp]
\psscalefirst
\centerline{\psfig{figure=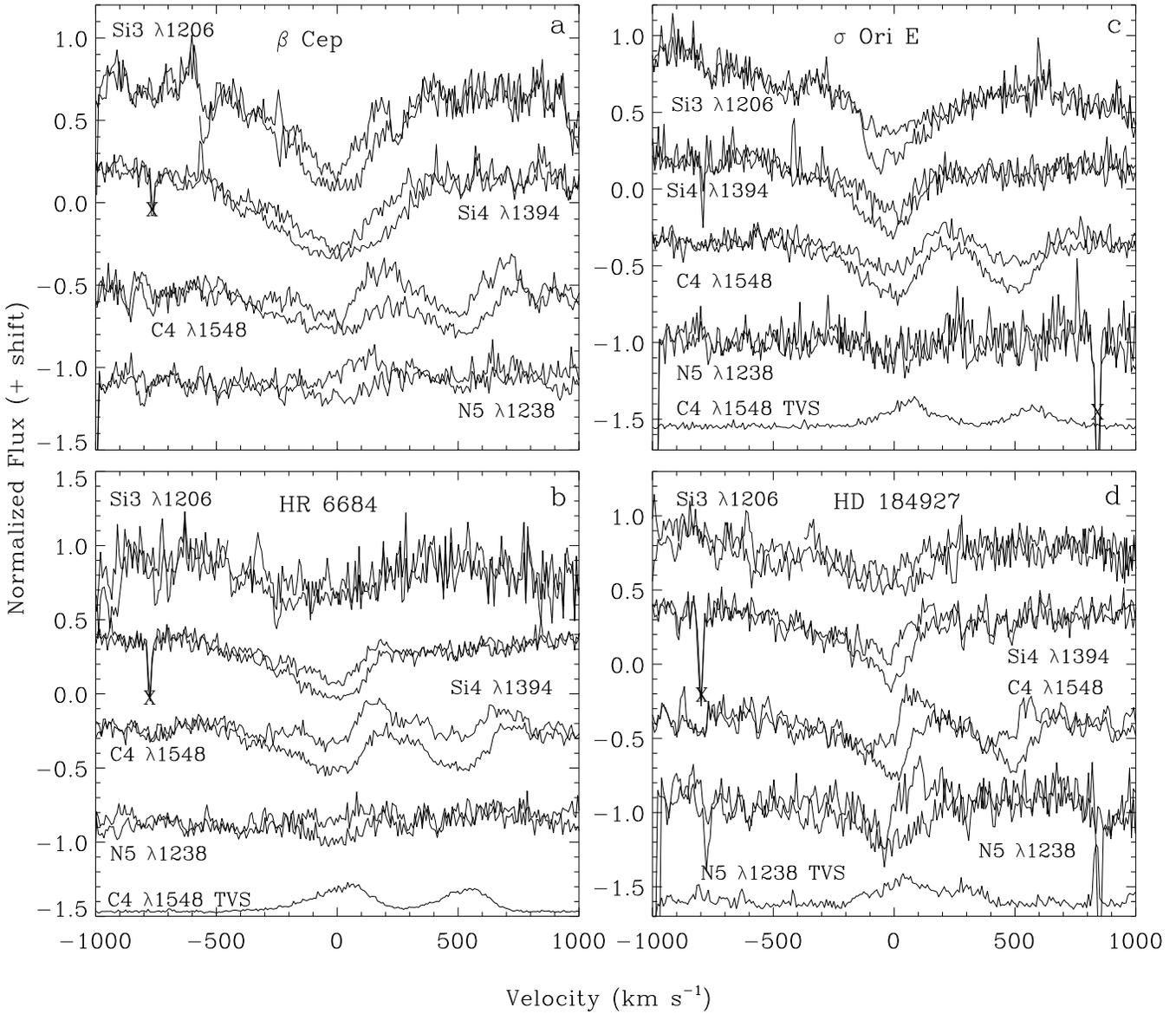,height=180mm,angle=90}}
\caption{{\bf a-d}
Comparison of spectra of Si\,III, Si\,IV, C\,IV, and N\,V
resonance lines of minimum (clear-star phase; dashed line) and maximum
(occultation phase, solid line) absorption strength (shifted for clarity).
Except for $\beta$\,Cep the normalized TVS spectrum is also given.
{\bf a}\, for $\beta$\,Cep, the corresponding spectra are from SWP 57573
(dashed line) and SWP 52514 (solid line),
{\bf b}\, for HR\,6684 the corresponding spectra are from SWP 55732 and
SWP 55526,
{\bf c}\, for $\sigma$\,Ori\,E the corresponding spectra are from
SWP\,07560 and occultation phase SWP\,15787,
{\bf d}\, corresponding spectra for HD\,184927 are from SWP 55652 and
SWP 55685.
}
\label{staremabs}
\end{figure*}

Fig.\,\ref{staremabs} depicts the total ranges in absorption and emission from
pairs of observations for the strong resonance lines for the program stars.
Inspection of these figures shows that profiles at clear-star (dashed lines)
phases differ from occultation phase profiles (solid lines) in the
following ways:
(1) the clear-star profiles of C\,IV and Si\,IV exhibit emissions in their
cores and red wings ($\S$\ref{resemis}),
(2) the Si\,III profiles exhibit narrower cores, and (3) the red and blue
far-wings of these lines are strong relative to photospheric line models.

   As with the weaker metallic lines,
one would hope to utilize the strengths of strong resonance lines
at clear-star phases as baselines to compare with the absorptions
at occultation phases. However, modeling the strong resonance lines
is a problem complicated by the presence of noncoherent scattering
(partial redistribution) in the line source function as well as
illumination of the torus by a diffuse radiation source (the jet).
In general, it is difficult to separate the effects of the torus from
both the photosphere and a jet. In particular, it is impossible in practice
to identify a reference photospheric profile from which to difference
absorption and emission of the torus-cloud
because the observed Si\,IV, C\,IV, and N\,V line profiles are
peculiar at all phases, exhibiting strong absorptions during occultations
and emissions at other times.
Nonetheless, points 2 and 3 above can provide hints of local
cloud conditions. For example, point \#2 suggests that the broad core of the
occultation-phase profiles is a signature of velocity broadening.
As to point \#3, the
extended, particularly red wings of the Si\,III and Si\,IV lines can best be
understood as enhanced strengthenings from a dense cloud. Consider that our
{\it CIRCUS} models for Si\,III $\lambda$1206 and the Si\,IV resonance doublet
indicate that densities less than 10$^{11}$ cm$^{-3}$ cannot be
distinguished from the clear-star (photospheric) profiles. Our best models require
cloud densities in the range of 10$^{12-13}$ cm$^{-3}$. Taking again the
simplest paradigm that the cloud temperature increases inward toward
the star, the wings of the strong silicon resonance lines suggest then
that cloud densities increase inwards too.

  Traces of the variations so readily apparent
in the Si\,IV, C\,IV, and N\,V lines are still present in Si\,III
$\lambda$1206 and even in the Al\,III doublet. However, the difference between
this line and the others is that the Si\,III line does not show any emission
at any phase. While this circumstance might lead to the hope that one can
model the $\lambda$1206 variations, we found in practice that we could not
reproduce the observed strengthening of this line without introducing a
substantial microturbulence. As an example, consider the Si\,III line in
$\beta$\,Cep. Table\,\ref{tab2}
shows a cloud-added absorption of 343 m\AA.\, {\it CIRCUS} models suggest a
maximum absorption for T$_{cloud}$ = 17\,000--19\,000\,K ($\S$\ref{siabs}).
Even for column densities of 3--10$\times$10$^{22}$ cm$^{-2}$ and a turbulence
of 20 km\,s$^{-1}$, our simulations
predict an absorption in the range 138--259 m\AA.
Low-temperature, pure-absorption ({\em non-LTE}) models also cannot reproduce
the observed strengthening. If one introduces a high turbulence of 50 
km\,s$^{-1}$ {\em and also} adopts a reasonable temperature and column density
(17\,000\,K and 3$\times$10$^{22}$ cm$^{-2}$), the observed absorptions
($\approx$340 m\AA),\, can then be reproduced. The Si\,III line's
sensitivity to turbulence
implied here is a consequence of the saturation in the core.
One may assume that the broad cores of the C\,IV and Si\,IV lines at 
occultation phases are also due to turbulence. As for Si\,III $\lambda$1206, 
moderately high turbulences and moderate temperatures 
produce good fits to the Al\,III doublet absorptions,
noted in Table\,\ref{tab3}. We also note similar good fits for HD\,184927.
This is not surprising in view of the similar
ionization potential of this ion: the Al\,III lines are probably formed in
nearly the same region of the cloud as Si\,III $\lambda$1206.

\subsubsection{The N\,V absorptions }

  The anomalously strong absorptions of the N\,V 1238\AA, 1242\AA\
doublet at occultation phases first drew our attention to the rotating
magnetic B stars. According to stellar atlases (e.g. Walborn et al.
\cite{walal2}), the spectra of {\em normal} B0--B2 V stars show no trace
of this doublet. Even for the hot star 10\,Lac
(O9; $T_{\rm eff}$ = 35\,400\,K),
we estimate that this line has a photospheric strength of only 175 m\AA~
(see the Brandt et al. \cite{braal1} atlas). Our {\it SYNSPEC} models reveal
that their presence is difficult to discern in photospheric spectra in stars
with $T_{\rm eff}$'s below about 28\,000\,K because of nearby line blends.
Grigsby \& Morrison (\cite{grimor}) have recently noted the anomalous
strengths of N\,V lines in spectra of the so-called {\em nitrogen
enhanced} late-O stars (Walborn et al. \cite{walal1}).
They attribute them to a photoionization of the wind by X-rays. However,
the wind explanation cannot apply to the rotating magnetic stars in our
sample because the lines do not show extended, high-velocity (blue) wings.
Moreover, the N\,V strengths in our program stars tend to be larger than
the features in these O-stars.
By elimination, we conclude that the N\,V absorptions
are formed in the circumstellar cloud(s).

  The N\,V absorptions in these stars are so large ($\approx$750 m\AA\, and
$\approx$530 m\AA\, over the {\it full} profile for $\beta$\,Cep and
HD\,184927, respectively) that they cannot be caused by enhanced
abundances or strong winds, or fit by our conventional strategies.
For example, our {\em standard} cloud models,
with both high cloud temperatures ($\sim$33\,000\,K) {\it and} a very
large high column density (10$^{23}$ cm$^{-2}$) can not reproduce them.
It is only when one considers models with both high cloud temperatures
and microturbulences together that the strong absorptions can be matched.
For such combinations the precise strength of the absorption is insensitive
to column length because the line is already so opaque.
To assess the effects of increasing both these parameters,
we carefully computed the underlying photospheric spectrum,
taking into account the star's $T_{\rm eff}$, log\,$g$, and $v$\,sin\,$i$.
The {\it CIRCUS} models showed that the N\,V line is extremely sensitive to
temperature, a fact that greatly restricts the range of variables in successful
models. For $\beta$\,Cep we settled on a
promising pair of models (normal abundances, column density of 10$^{22}$
cm$^{-2}$, microturbulence = 50 km\,s$^{-1}$). Then, a change from 28\,000\,K
to 29\,000\,K strengthens the equivalent width from 585 to 742 m\AA.\,
Increasing the microturbulence from 50 to 100 km\,s$^{-1}$ produces exactly
the same increase. Thus, both microturbulence and temperature are
sensitively constrained by the strengthening of this line.
We obtained
similar temperatures for $\sigma$\,Ori\,E and HR\,6684, with a turbulence
of 50 km\,s$^{-1}$. For HD\,184927 we were able to match the absorption
with a temperature of 30\,000--31\,000\,K, assuming a column density
of 1--2$\times$10$^{22}$ cm$^{-2}$ and 50 km\,s$^{-1}$. The best fit
models are summarized in Table\,\ref{tab4}.

\begin{figure}[tp]
\psscalefirst
\centerline{\psfig{figure=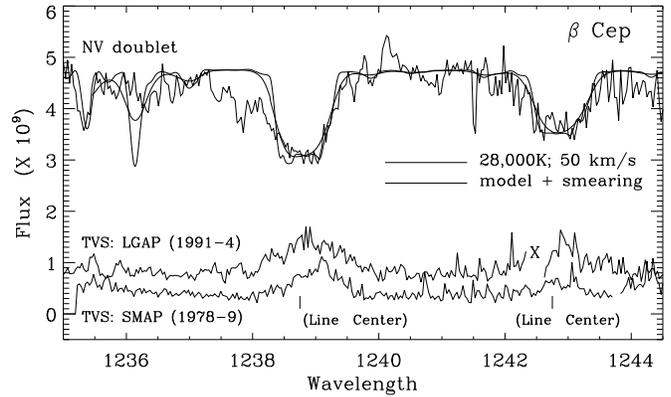,height=88mm,angle=90}}
\caption{
Comparison of mean spectrum of the N\,V doublet for $\beta$\,Cep compared to a
{\it CIRCUS} model fit (upper panel). The observations are taken from the
the occultation-phase spectra listed in Fig.\,\ref{starmetstvs}c. The fitted
model represents a cloud with the temperature and microturbulence shown.
The dashed fit is the result of smoothing with a a 100 km\,s$^{-1}$ gaussian.
Lower panel shows the TVS spectra divided by mean spectra for the ensemble
of large- (LGAP) and small-aperture (SMAP) observations of $\beta$\,Cep.
}
\label{fig6}
\end{figure}

  Fig.\,\ref{fig6} exhibits a model fit of the N\,V doublet for
$\beta$\,Cep at the occultation phase. The fitted
cloud absorption profile was computed with a {\it CIRCUS} model having a
coverage factor of 100\%, heated T$_{cloud}$ = 29\,000\,K,
a column density of 3.2$\times$10$^{22}$ cm$^{-2}$,
a microturbulence of 50 km\,s$^{-1}$, and a further convolution with a 100
km\,s$^{-1}$ gaussian. This macroturbulence-smearing was necessary to remove
the corners of a boxy-shape profile produced by microturbulence alone.
The parameters required to approximate the N\,V absorption of HD\,184927
are similar. Because HD\,184927 has a low surface metallicity, the required
cloud temperature is a little higher, $\approx$31\,000\,K, and the
derived micro- and macro-turbulences are then
50 km\,s$^{-1}$ and 100 km\,s$^{-1}$, respectively.

\section{Resonance line emissions}

\label{resemis}

  The spectroscopic hallmark of the rotating magnetic B stars is the strong
variation of their C\,IV and N\,V doublet lines with rotational phase,
ranging from strong absorption to redshifted emission. Since these
variations are not necessarily well ordered, only a few generalizations
can be made about their behavior. One overriding general characteristic is
that the emission and/or absorption components are always present to some
degree. A second characteristic is that although
small differences can arise in same-phase profiles of different epochs,
the general variations are remarkably repeatable over at least several years.
For example, a narrow redshifted emission feature seems to be a common
feature in the 1238\,\AA\, profiles of HD\,184927 at phases near 0.85 and 0.15,
i.e. at the beginning and end of the range of clear-star phases.
For some stars, such as HR\,6684, the N\,V profiles almost always show emission
components, even when much of the profile is strongly in absorption. For other
stars like $\sigma$\,Ori\,E, the N\,V emissions are difficult to detect at all.
The weakness of N\,V emissions in $\sigma$\,Ori\,E,
along with the weakness of Si\,IV emissions for all of the program
stars, probably explains why these emissions have gone unreported
for so long in studies of He-strong stars.

  All four of our stars show detectable emissions in lines of at least two
ions. In two of our four program
stars, the N\,V lines exhibit the strongest emissions, while for
HR\,6684 or $\sigma$\,Ori\,E it is the C\,IV emissions which are the strongest.

 In some cases, the N\,V emissions of these stars are almost completely
confined to the red side of the profile, and indeed in all cases the TVS
is larger there. The lower portions of Figs.\,\ref{staremabs} b, c, d
and \ref{fig6}, depict the TVS spectrum divided by the mean
profile for the C\,IV profiles or N\,V profiles.
We depict the ratio of the TVS to the mean spectrum
in order to account for absorptions of the strong line core which would
otherwise suppress fluctuations at low velocities. The TVS-ratio
profiles generally peak at +60--125 km\,s$^{-1}$, and their red wings
extend to +300--400 km\,s$^{-1}$ in each star with remarkable uniformity.
Thus, the edge velocities are not related to the peak emission
strength, the star's metallicity, pulsational status, or $v\,sin\,{\it i}$.

   A comparison of the TVS-ratio spectrum for $\beta$\,Cep among both
large- and small-aperture observations in
Fig.\,\ref{fig6} gives a qualitative idea of the robustness and
constancy of the activity spectrum over even widely separated epochs for
this star. It is interesting that the TVS profiles for C\,IV and N\,V
do not exhibit a tail extending to the far blue wing. This fact
shows that the stars' winds do not vary much over stellar longitude.
In all, one finds that the redshifted emissions are uncorrelated
with wind visibility in the blue wings of the lines.
Likewise, the emissions do not appear to arise from
the wind, so one must look to a different mechanism to explain them too.

   We can quantify the temperature range required to reproduce the
N\,V emissions by again using {\it CIRCUS} models, again taking
$\beta$\,Cep as an example. As one increases the cloud plasma temperature
in these models, the N\,V lines first show increased
absorptions at temperatures. These absorptions peak in the range
$\sim$35\,000--40\,000\,K, both for LTE and non-LTE models. As the
temperature increases the line goes into emission.
One may place interesting upper limits on the line formation region of
the N\,V emissions because at temperatures in excess of
40\,000\,K these regions of the cloud would contribute
an unacceptably large fraction to the combined continuum
flux of the star and cloud.
Such models would produce a nearly
sinuosoidal modulation in continuous flux as the torus co-rotates around
the star. This is not indicated by our continuum light curve data.
so we can conclude
that any continuum flux from a bright emission patch near the poles
is negligible. An additional consequence of these arguments is that the N\,V
line emissions are optically thin.
This result is also supported
by the high ratio (1.5--2) of the relative r.m.s. amplitudes of the two
lines in Fig.\,\ref{fig6} (the doublet components have a gf ratio of 2).

  In assuming the optical thinness condition and an areal coverage
factor of unity, it becomes straightforward to model the $\lambda$1238
emission equivalent width. Again taking $\beta$\,Cep as an example,
the observed emission over the full profile is $\approx$ -220 m\AA.\, In the
models there is only a small trade-off between gas temperature and column
density, e.g., a choice of parameters 49\,000\,K and
3$\times$10$^{17}$ cm$^{-2}$
or 50\,000\,K and 10$^{18}$ cm$^{-2}$; the limiting condition for
$\tau_{\lambda1238}$ = 1.
Because of the sensitivity of N\,V ionization with temperature,
the results for HD\,184927 (full profile equivalent width $\sim$ -0.055\,\AA)
are similar to those for $\beta$\,Cep: we find an optimal temperature
of 45\,000\,K for the resulting length of the N\,V emission region is
1$\times$10$^{11}$ cm$^{-3}$,
the N\,V-emitting region comes out to be prohibitively
thin, only $\sim$100 km. This is roughly
100$\times$ smaller than a scale height in the torus.
The shock depth can be reconciled to the expected range of values if the
assumed coverage factor for N\,V emission is $\le$1\% rather than
the initially assumed value of 100\%.

\section{Interpretation}
\label{interp}

  A key result of the previous section is that the UV line absorptions of
a rotating magnetic B star vary because of a cloud passing in front
of the star. Thus, there is no need to invoke (nor evidence for)
an inhomogeneous metallic composition on the surfaces of these stars.
The observations also suggest that the mass in these clouds is nearly
constant over time. To maintain a constant mass,
the wind accreted by the cloud must somehow settle back to the star
and/or escape through its outer point. Indeed, perhaps it is the settling of
cloud material onto the surface that is responsible for the high densities
we found in the {\em inner cloud regions}. As a consequence of
this recirculation, the cloud has virtually the
same metallic composition as the star's polar caps and
indeed the rest of its surface.

  The numerical analysis of line strength variations in $\S$\ref{analy}
indicates that the cloud line components form in an extended
environment of a column density of roughly 1$\times$10$^{23}$ cm$^{-2}$.
In the detailed models of the cloud spectra of HD\,184927 and $\beta$\,Cep,
the lines are formed over a range of temperatures consistent with the stellar
photosphere's radiation field. Although there is no evidence for
long-term changes in cloud conditions, there is a strong suggestion of
small- and large-scale turbulence in the inner regions. The strong
wings of the Si\,IV and Si\,III resonance lines argue for increasing cloud
densities in these regions, which seem to approach photospheric values
at the inner edge of the torus.

  The redshifted emissions in the N\,V and C\,IV lines observed during the
magnetic pole-on phases demonstrate that downflows occur in the
vicinity of the rotating, magnetic B-stars.
In our picture they are most likely to arise from shocks because of the 
rapid deceleration of the wind particles as they impact the pre-existing
torus (cf. Babel \& Montmerle \cite{babmon1}). This produces
a redshift during the phase when the magnetic pole is visible because the
wind has followed the curved lines of force and is then moving away from
the observer. However, additional possibilities are possible. One should
also consider the possibility of a stalled
wind which impacts the star's surface near its point of origin. (In principle,
it is also possible that magnetohydrodynamical processes, or Kelvin-Helmholtz
instabilities; see Shore et al.
(\cite{shoal0}), operate, but we see no need for this in view of
of the potential of steady-state hydrodynamical iteractions alone.)

   In examining the evidence, we have
found two emission characteristics that cast doubt on whether the emissions
are produced by reaccretion, and we review them in the following.
We discovered the first of these characteristics by recalling that SB90
had drawn attention to {\em red-shaded} C\,IV emissions in two He-strong
stars for
which periods are not known, HD\,96446 and HD\,58260. SB90 also
noted that these features cannot arise from spherically symmetric shells,
i.e. they are not true P-Cygni profiles. The SB90 star HD\,96446 is likely to
meet our criteria for membership among the rotating magnetic B stars because
it also shows N\,V line emission. In addition, its emissions are
variable. The case of HD\,58260 is particularly interesting in this
analysis for two reasons. First, SB90 have pointed out
that we probably observe this star at a pole-on aspect
(the magnetic and rotational poles are nearly coincident), so its emissions
do not vary. Second, this star's C\,IV line emissions are both optically
thick and strong -- indeed, it attains a level of 0.5I$_{cont}$.
In attempting to model C\,IV emissions, we found that it was not possible
to reproduce this emission strength within the context of our models.
Even our {\em best} model parameters
(T$_{cloud}$ = 30\,000\,K, turbulence = 50--100 km\,s$^{-1}$, high
column densities, coverage factor = 1) produce C\,IV emissions of
only 0.20 I$_{cont}$. We also found that the efficiency for producing
the emission peaks very sharply at 30\,000\,K, so areal coverage factors of
at least 3--4$\times$ are probably necessary.
{\it Since this factor is significantly greater than one, the emissions
cannot be produced just on the stellar surface.} The area of the cloud-torus
is more than adequate to produce the total emission, so we are led to the
conclusion that the emissions are caused at the wind-cloud
interface.

  The second characteristic we used to decide between the competing
shock models for line emission came from the detailed examination of the
evolution of the C\,IV and N\,V emissions in both $\beta$\,Cep and
$\sigma$\,Ori\,E.
We discovered that an emission component appears
first at the end of the occultation phase.
In a model in which the downflows fall onto the star's surface, we would
expect the line-emitting material to share the star's rotation because of
its co-rotation with the star. As the cloud clears the receding limb of the
star at the end of its occultation, emission is produced as the failed-wind
stream shocks at a magnetic pole (now appearing at the opposite limb of the
star). The emissions from this region will have an algebraically smaller
radial velocity both because of the star's rotation and
because the projected component of the infall velocity at this time is
nearly zero. The emissions should then migrate redward smoothly across the
line profile in this model as both velocities become more positive.
The actual observations indicate that the emission component
first appears weakly at a common point in the line
profiles of both stars ($\sim$-200 km\,s$^{-1}$) and then migrates to the red.
This smooth behavior is suddenly interrupted by the strong appearance of
a red emission component at +0--100 km\,s$^{-1}$. This particular evolution
is difficult to understand as a consequence of shocks occurring on the
stellar surface,
particularly for $\beta$\,Cep and $\sigma$\,Ori\,E, which have very
different rotational velocities. In contrast, the disk interpretation is
almost {\em too easy}, suffering only from an embarrassment of several
nonunique geometrical interpretations.

  A related question is how the N\,V absorptions are formed during occultation
phases, sometimes even coexisting with the emissions. The hypothesis that
the N\,V emissions are emitted from the torus interface provides an advantage
of restricting the formation of the emission to the same general site as
the absorption. Both sites require a high temperature. Unlike the emission
component, the absorption is optically thick. {\it CIRCUS} models show that
absorption of N\,V is necessarily formed in a turbulent medium having a warm
temperature of about 30\,000\,K. These results suggest that the N\,V absorption
is formed in a large region of the cloud in which C\,IV and N\,V are
overionized. We suggest that the wind-cloud shocks generate
this overionization.

  Still, it is not clear whether the heating required to produce the N\,V
absorption occurs from a conversion of mechanical or radiative energy.
The problem with either mechanism is that it is difficult to heat a large
enough cloud volume to produce the amount of absorption observed.
For example, turbulence can circulate shock-heated gas but probably only
to a few scale heights. For representative cloud parameters
(30\,000\,K, 10$^{11}$ cm$^{-3}$) this length is probably only
10$^{5}$ km or less. Radiative models encounter similar difficulties if the
radiation from the shocks is emitted primarily in the EUV.
The photoionization radius for N$^{4+}$ in our cloud models
is only $\sim$10$^{3}$ km. In contrast, {\it CIRCUS} models
that match the N\,V absorptions require a full areal coverage
factor, so the scale length for energy dissipation must be at least
1R$_{*}$. The radiative model can be saved if the
shocks generate soft X-rays (1/2--1 keV) which
would be transparent to the cloud over its full length. Such irradiation
would
produce a broad distribution of ion states (Cohen et al. \cite{cohal2}) and
would
thus account for the observed strong absorptions produced by the cloud over
a range of ion stages. The evidence for X-ray heating on these stars is
ambiguous. Although the mean X-ray levels are known to be low,
little is known yet about variability. The kinetic model in turn
might be reconciled with models of relativistic particles penetrating
to great depths into the torus of $\sigma$\,Ori\,E. Alternatively, it
may be that porosity, rarefied zones, and/or Alfv\'en waves play roles
in transporting mechanical energy into the cloud. Although we favor the
radiative model, all of these possibilities
are speculative. Thus, the question of the origin of the N\,V
absorptions remains open.

\section{Conclusions}
\label{conclu}

  The present work confirms the finding by Henrichs et al. (\cite{henal1})
that a separate class of rotating magnetic stars exists among the early-type
main sequence B stars. Taking $\beta$\,Cep as a case in point, one finds that
a dipolar field of only a few hundred gauss can suffice to guide a polar
wind from the surface to its equatorial regions via closed magnetic loops.
The wind particles are abruptly halted near the magnetic equator where they
establish a quasi-static, torus-shape structure (SB90).
Our picture differs from the picture of GH97 and other investigators
in two important respects. First, we have found that variations of
the UV and probably optical lines (other than of helium) are
due to absorptions in the torus-cloud.
Second, we have remarked that the blue wings of the Si\,III, Si\,IV, and
C\,IV resonance lines remain remarkably constant with rotational phase
for all the program stars.
Let us examine a few ramifications of these results.

  If indeed the variable line absorptions are caused by the clouds, and if
the metallic composition is nearly constant across their surfaces, their
low-metallicities all over the surface will be maintained only if cloud
material settles back to the star more efficiently than its wind streams
upward. One may speculate
as to how this might happen.
In our model, the result of this settling is that the inner
torus has a high density and thus controls the configuration of the local
field lines, rather than the other way round. For example, the weight
of the particles over the equator may cause the field lines to sag,
as in some dense solar quiescent prominences, and
particles could return directly to the surface.
Turbulence may also help recirculate cloud particles to the star.

  The phase-independent nature of the wind is more difficult to understand.
GH97 pointed to the two types of wind, one originating from the surface
(magnetic pole), the second, an exo-magnetospheric wind
originating at the outer edge of the torus. If the absorption contributions
of the two wind components were equal in the blue wings, one would
expect the fluxes of their blue wings to vary sinusoidally
twice per cycle -- yet this is not observed. However, even if the wind mass
fluxes were comparable, it is likely that their observable contributions
in the blue wings of the line profiles could still be {\it unequal} for a
few reasons. For example,
the exo-magnetospheric wind will
be more likely to remain visible over a wide range of phases, both as a
consequence of its inverse-square density dependence with distance and
because its streamlines will arc {\em backwards} to maintain angular
momentum conservation. These curved trajectories will cross our lines of
sight over a variety of rotational phases as they go off to infinity,
diluting any modulation of wind absorption at any particular phase.
The key requirement of
all such arguments is that the wind remains visible over long path lengths.

  As outlined in $\S$\ref{sketch}, a necessary requisite of our picture of
rotating magnetic B stars is that closed field loops divert the wind flow
from the magnetic poles to the torus at the magnetic equator. In this picture
stalled wind particles return to the magnetic caps and establish a large
composition gradient. This does not occur in nonmagnetic stars because
envelope mixing destroys composition gradients before they can be established.
Also, similar processes may take place in hot Bp analogs (e.g.,
$\theta$$^{1}$\,Ori\,C), but its wind is so strong that chemical separation
cannot be expected to take place. The results of this paper suggest
indirectly that composition gradients are destroyed
in magnetic, pulsating stars, such as in $\beta$\,Cep and HR\,6684. Yet, there
are fundamental puzzles in even this simple picture. For example, what is
to be concluded about HR\,6684, which has low-metal but normal-He abundances?
Perhaps the metals leave this star in the wind while both hydrogen and helium
atoms are reaccreted. Such circumstances are envisioned by Hunger \& Groote
(\cite{hungro2}), but only in late-type B stars. A second
puzzle is the Be-like episodes exhibited by $\beta$\,Cep
perhaps every 30 years (Kaper \& Mathias \cite{kapmat}, Panko \& Tarasov
\cite{panko}). The tight scatter in the resonance line absorptions in
Fig.\,\ref{starewlines}a and in Fig.\,5 of Henrichs et al. (\cite{henal2})
suggests that the presence or absence of a flattened Be disk does
not affect torus-cloud signatures noticeably. Thus, if these ejections are
related to the Be-phenomenon, the UV phenomenology of the two types of
circumstellar structures is vastly different (in addition to their physical
morphologies).

  In $\S$1 and $\S$5 we listed 13 probable members of the rotating magnetic
B-star group (including $\theta$${^1}$\,Ori\,C).
Rotating magnetic B stars
appear also to include an indeterminate number of early-B main sequence stars
with conditions that promote envelope mixing (pulsation, rapid rotation).
One such example may be the unusual star HD\,144941 (see also GH97).
This star lies within the region of the H-R Diagram populated by the
He-strong stars and shows extreme He-richness and metal
deficiencies. Its high galactic latitude, and high radial velocity (-52.6
km\,s$^{-1}$) led Harrison \& Jeffery
(\cite{harjef1}, \cite{harjef2}) and others to consider it as a member of
the hydrogen-deficient B star class.
Moreover, Harrison \& Jeffery (\cite{harjef2}) found that
the Fe abundance is just as deficient as the CNO elements and suggested that
these abundances are primordial. To test this result, we inspected two
high-dispersion {\it IUE} spectra obtained for this star reveal redshifted
emissions in both the C\,IV and Si\,IV lines as well as a strong wind. These
spectra show anomalously strong N\,V absorption
so perhaps HD\,144941 is a member too.
We must also remark that the search for magnetic and 
rotational periods is very labor and telescope intensive, so 
new members of this class might be found in the long run by examining 
spectra of early-B stars for the presence of anomalous N\,V absorptions 
and/or Si\,IV, C\,IV, N\,V emissions. 

  The physical attributes of the tori around the rotating magnetic stars
can also be used to clarify our understanding of {\em classical} stars. 
Several studies (see Balona \cite{bal}) have suggested that co-rotating clouds 
are responsible for periodic spectroscopic and/or photometric varations
in these stars. The difficulty in distinguishing between this explanation
and nonradial pulsation is that optically thick clouds can mimic cool regions 
on the atmosphere. We are optimistic that some of the results in this paper 
that indicate the presence of redshifted components in resonance 
lines, the selective variation of low-excitation lines, and high turbulence 
together can help to discriminate between
models responsible for the variability of these stars.

  A general understanding of physical conditions leading to co-rotating 
magnetospheres will be aided by future studies of the magnetic rotating B 
variables (with the probable extension to the {sn} He-weak stars (see
Shore et al. \cite{shoal0}, \cite{shoal1}), hot He-strong 
analogues like $\theta$$^{1}$\,Ori\,C, Ap stars like IQ\,Aur (Babel \& 
Montmerle \cite{babmon3}), magnetic Herbig Ae-Be stars, 
and late-type active stars such as AB\,Dor.
The results of this paper point to a number of desiderata which 
will clarify the nature of the class of rotating magnetic B stars 
in particular and also contribute to its global understanding.
We close by listing a few of them:

\begin{itemize}

  \item the {\it IUE} archive can be conveniently investigated to search for
   additional examples of early-B type stars with anomalous N\,V absorptions
   and redshifted emissions among its C\,IV and N\,V resonance lines,

  \item far-UV {\em snapshot} observations can be undertaken to determine 
   whether the anomalous absorptions and redshifted emissions extend to 
   the O\,VI doublet, 

  \item the X-ray fluxes from these stars can be monitored around their
   rotational cycles to shed some light on the production mechanism of 
   the C\,IV, N\,V absorptions,

   \item optical observations of the high Balmer lines can be carried out
   over rotational phase among all the candidate members of this class
   with a view of establishing whether cloud density varies with time
   and as a function of relevant stellar parameters,

   \item spectral imaging can profitably be carried out in the H$\alpha$
   line and ultimately in the strong ultraviolet resonance lines from 
   space-borne interferometers.

\end{itemize}

{\em Acknowledgments.}
We wish to thank the {\em invisible author,} Dr. Ivan Hubeny, for his 
making his suite of codes available for public access and for answering
a number of questions we had about their efficient use. We also acknowledge 
an {\it IDL} wrapper for {\it CIRCUS} written by Dr. Richard Robinson for
another study that permitted us to exploit the flexibility of the Hubeny codes. 
It is our pleasure to thank Dr. Huib Henrichs for providing us with a 
preprint of recent Zeeman analysis work on $\beta$\,Cep. We are grateful to 
Dr. Kurt Hunger, Drs. Steve Cranmer and David Bohlender for important 
discussions. We also thank Drs. Luis Balona and Kurt Hunger for a critical 
reading of the manuscript. We appreciate a number of very helpful comments
made by the referee, Dr. Steven Shore. This work was largely supported
through NASA contracts NASA NAG 5-6733, 5-8793 and GO-6086.01-94A.

\newpage

\end{document}